\documentclass{article}
\usepackage[utf8x]{inputenc}
\usepackage{agda}
\usepackage[nocenter]{qtree}
\usepackage{amssymb,amsthm,amsmath}
\usepackage{fullpage}
\usepackage{url}
\usepackage{tikz}
\usepackage{tikz-cd}
\usepackage{listings}
\usepackage{multicol}
\usepackage{stmaryrd}
\usepackage{proof}
\usetikzlibrary{quotes,fit,positioning}
\tikzset{>=latex}

\newtheorem{theorem}{Theorem}[section]

\newtheorem{definition}[theorem]{Definition}
\newtheorem{proposition}[theorem]{Proposition}

\newcommand{\inkscape}[2][1.5]{
\begin{center}
\scalebox{#1}{
\includegraphics{inkscape/#2}
}
\end{center}
}

\newcommand{\presumtype}{\uplus}
\newcommand{\preprodtype}{*}
\newcommand{\sumtype}{+}
\newcommand{\prodtype}{\times}

\newcommand{\inleft}[1]{\textsf{left}~#1}
\newcommand{\inright}[1]{\textsf{right}~#1}

\newcommand{\identlp}{\ensuremath{\mathit{unite}_{\sumtype}\mathit{l}}}
\newcommand{\identrp}{\ensuremath{\mathit{uniti}_{\sumtype}\mathit{l}}}
\newcommand{\identlsp}{\ensuremath{\mathit{unite}_{\sumtype}\mathit{r}}}
\newcommand{\identrsp}{\ensuremath{\mathit{uniti}_{\sumtype}\mathit{r}}}
\newcommand{\swapp}{\ensuremath{\mathit{swap}_{\sumtype}}}
\newcommand{\assoclp}{\ensuremath{\mathit{assocl}_{\sumtype}}}
\newcommand{\assocrp}{\ensuremath{\mathit{assocr}_{\sumtype}}}
\newcommand{\identlt}{\ensuremath{\mathit{unite}_{\prodtype}\mathit{l}}}
\newcommand{\identrt}{\ensuremath{\mathit{uniti}_{\prodtype}\mathit{l}}}
\newcommand{\identlst}{\ensuremath{\mathit{unite}_{\prodtype}\mathit{r}}}
\newcommand{\identrst}{\ensuremath{\mathit{uniti}_{\prodtype}\mathit{r}}}
\newcommand{\swapt}{\ensuremath{\mathit{swap}_{\prodtype}}}
\newcommand{\assoclt}{\ensuremath{\mathit{assocl}_{\prodtype}}}
\newcommand{\assocrt}{\ensuremath{\mathit{assocr}_{\prodtype}}}
\newcommand{\absorbr}{\ensuremath{\mathit{absorbr}}}
\newcommand{\absorbl}{\ensuremath{\mathit{absorbl}}}
\newcommand{\factorzr}{\ensuremath{\mathit{factorzr}}}
\newcommand{\factorzl}{\ensuremath{\mathit{factorzl}}}
\newcommand{\dist}{\ensuremath{\mathit{dist}}}
\newcommand{\factor}{\ensuremath{\mathit{factor}}}
\newcommand{\distl}{\ensuremath{\mathit{distl}}}
\newcommand{\factorl}{\ensuremath{\mathit{factorl}}}
\newcommand{\iso}{\leftrightarrow}
\newcommand{\proves}{\vdash}
\newcommand{\idc}{\mathit{id}\!\!\leftrightarrow}
\newcommand{\Rule}[4]{
\makebox{{\rm #1}
$\displaystyle
\frac{\begin{array}{l}#2 \\\end{array}}
{\begin{array}{l}#3      \\\end{array}}$
 #4}}
\newcommand{\jdg}[3]{#2 \proves_{#1} #3}
\newcommand{\sem}[1]{\ensuremath{\llbracket{#1}\rrbracket}}

\DeclareUnicodeCharacter{9678}{\ensuremath{\odot}}
\DeclareUnicodeCharacter{9636}{\ensuremath{\Box}}
\DeclareUnicodeCharacter{10231}{\ensuremath{\leftrightarrow}}

\tikzstyle{func}=[rectangle,draw,fill=black!20,minimum size=1.9em,
  text width=2.4em, text centered]

\title{Embracing the Laws of Physics: \\ Three Reversible Models of Computation}
\author{Jacques Carette \qquad\qquad Roshan P. James \qquad\qquad Amr Sabry \\
McMaster University \qquad\qquad Google \qquad\qquad Indiana University}

\newtheorem{thm}{Theorem}[section]
\newtheorem{defn}{Definition}[section]
\newtheorem{prop}{Proposition}[section]

\newcommand{\fin}[1]{\ensuremath{\left[#1\right]}}
\newcommand{\Nat}{\ensuremath{\mathbb{N}}}
\newcommand{\true}{\mathit{true}}
\newcommand{\false}{\mathit{false}}
\newcommand{\Gpd}{\ensuremath{\mathsf{Groupoid}}}

\begin{document}
\maketitle

\begin{abstract}
  Our main models of computation (the Turing Machine and the RAM) and
  most modern computer architectures make fundamental assumptions
  about which primitive operations are realizable on a physical
  computing device. The consensus is that these primitive operations
  include logical operations like conjunction, disjunction and
  negation, as well as reading and writing to a large collection of
  memory locations. This perspective conforms to a macro-level view of
  physics and indeed these operations are realizable using macro-level
  devices involving thousands of electrons. This point of view is
  however incompatible with computation realized using quantum devices
  or analyzed using elementary thermodynamics as both these
  fundamental physical theories imply that information is a conserved
  quantity of physical processes and hence of primitive computational
  operations.

  Our aim is to re-develop foundational computational models in a way
  that embraces the principle of conservation of information. We first
  define what information is and what its conservation means in a
  computational setting. We emphasize the idea that computations must
  be reversible transformations on data. One can think of data as
  modeled using topological spaces and programs as modeled by
  reversible deformations of these spaces. We then illustrate this
  idea using three notions of data and their associated reversible
  computational models. The first instance only assumes unstructured
  finite data, i.e., discrete topological spaces. The corresponding
  notion of reversible computation is that of permutations. We show
  how this simple model subsumes conventional computations on finite
  sets. We then consider a modern structured notion of data based on
  the Curry-Howard correspondence between logic and type theory. We
  develop the corresponding notion of reversible deformations using a
  sound and complete programming language for witnessing type
  isomorphisms and proof terms for commutative semirings.  We then
  ``move up a level'' to examine spaces that treat programs as data,
  which is a crucial notion for any universal model of computation. To
  derive the corresponding notion of reversible programs between
  programs, i.e., reversible program equivalences, we look at the
  ``higher dimensional'' analog to commutative semirings: symmetric
  rig groupoids. The coherence laws for these groupoids turn out to be
  exactly the sound and complete reversible program equivalences we
  seek.

  We conclude with some possible generalizations inspired by homotopy
  type theory and survey several open directions for further research.

\end{abstract}

\section{Reversibility, the Missing Principle}

What kind of operations can computers perform? This question has been
answered several times in the last hundred years, where each answer
proposes an abstract \emph{model of computation} that specifies
allowable operations and (usually) their cost. The emerging consensus,
reflected in both early models of computations such as the Turing
Machine and the RAM as well as in the early Von Neumann models and in
modern computer architectures, is that basic computer operations
include logical operations like conjunction, disjunction, and
negation, as well as reading from and writing to a large (infinite)
collection of memory locations. From this small set of primitive
operations emerges all higher-level programming languages and
abstractions.

No doubt, this consensus on the available primitive physical
operations has been successful. Futhermore, these operations
\emph{can} indeed be performed on a computer. Yet, today, with a
possible quantum computing revolution in sight and an unprecedented
explosion in embedded computers and cyber-physical systems, there are
reasons to re-think this foundational question. In fact, the
calls to re-think this foundational question have been proclaimed by
physicists almost forty years ago:

\begin{quote}
  \textbf{Toffoli 1980~\cite{Toffoli:1980}:} Mathematical models of
  computation are abstract constructions, by their nature unfettered
  by physical laws. However, if these models are to give indications
  that are relevant to concrete computing, they must somehow capture,
  albeit in a selective and stylized way, certain general physical
  restrictions to which all concrete computing processes are
  subjected.

  \textbf{Feynman 1982~\cite{springerlink:10.1007/BF02650179}:}
  Another thing that has been suggested early was that natural laws
  are reversible, but that computer rules are not. But this turned out
  to be false; the computer rules can be reversible, and it has been a
  very, very useful thing to notice and to discover that. This is a
  place where the relationship of physics and computation has turned
  itself the other way and told us something about the possibilities
  of computation. So this is an interesting subject because it tells
  us something about computer rules.
\end{quote}

\noindent These quotes by Toffoli and Feynman both highlight the
consequences of two obvious observations: (i) all the operations that
a computer performs reduce to basic physical operations; and (ii)
there is a mismatch between the logical operations of a typical model
of computation (which are logically irreversible) and the fundamental
laws of physics (which are reversible). One could certainly dismiss
the mismatch as irrelevant to the practice of computing but our thesis
is that the next computing revolution is likely to be founded on
revised models of computation that are designed to be in closer harmony
with the laws of physics.

After a detailed introduction on the origins of \emph{logically
reversibile computer operations} and an excursion into the origins of
\emph{irreversible computer operations}, we will develop in detail three
reversible models of computation and discuss their potential
applications.

\paragraph*{Maxwell's Daemon.}
To fully appreciate the missing principle of \emph{reversibility} in
conventional computing, we go back to an old thought experiment by
J. C. Maxwell. The details are codified in a letter that Maxwell wrote
to P. G Tait in 1867 -- the letter, whose ideas are now known as
\emph{Maxwell's Daemon}, tells of a thought experiment that seems to
indicate that intelligent beings can somehow violate the second law of
thermodynamics, thereby violating physics itself. Many resolutions
were offered for this conundrum (for a compilation, see the book by
Leff and Rex~\cite{leff1990}), but none withstood careful scrutiny
until the establishment of \emph{Landauer's Principle} in
1961~\cite{Landauer:1961} -- a principle whose experimental validation
happened in 2012~\cite{berut2012experimental}.

Maxwell's Daemon appears to violate the second law of thermodynamics by
having a tiny ``intelligence'' observing the movement of individual
particles of a gas and separating fast moving particles from slow
moving ones, thereby reducing the total entropy of the
system. Landauer's resolution of the daemon relied on two ideas that
had taken root only a few decades earlier: the formal notion of computation
(through the work of Turing~\cite{turing}, Church~\cite{church51}, and
others) and the formal notion of information (through the work of
Shannon~\cite{shannon1948}). Landauer reasoned that the computation
done by the finite brain of the daemon involves getting information
about the movement of molecules, storing that information, analyzing
that information to act on it, and then --- and this is the critical
step --- overwriting it to make room for the next computation.  In
other words, the computation that is manipulating information in the
daemon's brain \textit{must be thermodynamic work}, thereby bringing
the daemon back into the fold of physics.

This is a strange and wonderful idea: information, physics, and
computation are inextricably linked. In contrast, when the early
models of computation were developed, there was no compelling reason
to take the information content of computations into consideration
-- in fact, at that time there was no quantifiable notion of
information. These models followed in the footsteps of logic where,
following hundreds of years of tradition, the truth of a statement was
seen as \emph{absolute} and independent of any reasoning, understanding,
or action. Statements were either true or false with no regard to any
\emph{observer} and the idea that statements had information content that
should be preserved was outside the classical understanding of
logic. Hence the fact that conventional logic operations such as
conjunction and disjunction were logically irreversible and hence lose
information was not a concern. Landauer's observation implied however
that ideas in each field have consequences for the
other~\cite{bennett:1973:lrc,bennett1985fundamental,bennett2010notes,bennett2003notes,Baker:1992:NFT,baez2011physics,dblp:conf/csfw/malacarias12}. To
really appreciate this fact, we delve deeper into the origin of our
computational models and argue that they are essentially reflections
of contemporary laws of physics.

\paragraph*{Origins of Computational Models.}
Current high-level programming languages as well as current hardware
are both based on the mathematical formalization of logic developed by
De Morgan, Venn, Boole, and Peirce in the mid to late 1800s. Going
back to Boole's 1853 book entitled \textit{An Investigation of the Laws
  of Thought, on which are Founded the Mathematical Theories of Logic
  and Probabilities}, we find that the opening sentence of Ch.~1 is:
\begin{quote}
  The design of the following treatise is to investigate the fundamental laws
  of those operations of the mind by which reasoning is performed;
\end{quote}
which clearly identifies the \emph{source} of the logical laws as
mirroring Boole's understanding of human reasoning. A few chapters
later, we find:
\begin{quote}
  \textbf{Proposition IV.}  That axiom of metaphysicians which is termed the
  principle of contradiction, and which affirms that it is impossible for any
  being to possess a quality, and at the same time not to possess it, is a
  consequence of the fundamental law of thought, whose expression is 
  $x^2 = x$.
\end{quote}
This ``law'' is reasonable in a classical world but is violated by the
postulates of quantum mechanics. Although a detailed historical
analysis of Boole's ideas in the light of modern physics is beyond our
scope, the above quotes should convey the idea that our elementary
computing notions date back to ideas that were thought reasonable in
the late 1800s.

Machines that ``compute'' are quite old. M\"{u}ller (1786) first
conceived of the idea of a ``difference machine,'' which Babbage
(1819--1822) was able to construct. There are other computer
precursors as well -- the first stored programs were actually for looms, most
notably those of Bouchon (1725) which were controlled by a paper
tape, and of Jacquard (1804), controlled by chains of punched cards.
But it was only in the 20th century that computer science emerged as a
formal discipline. One of the pioneering works was Alan Turing's
seminal paper~\cite{turing} of 1936 which established the idea that
computation has a formal interpretation and that all computability can
be captured within a formal system. Implicit in this achievement
however is the idea that abstract models of computation are just that
-- \emph{abstractions of computation realized in the physical world.}
Indeed, going back to Turing's 1936 article \textit{On Computable
  Numbers, with an Application to the Entscheidungsproblem,} the
opening sentence of Sec. 1 is:
\begin{quote}
  We have said that the computable numbers are those whose decimals
  are calculable by finite means [\ldots] the justification lies in
  the fact that the human memory is necessarily limited.
\end{quote}
In Sec. 9, we find:
\begin{quote}
  I think it is reasonable to suppose that they can only be squares
  whose distance from the closest of the immediately previously
  observed squares does not exceed a certain fixed amount.
\end{quote}
It is worth noting that these assumptions are both physical (on
distances) and metaphysical (on restrictions of the mind).  If we take
the human mind to be a physical ``machine'' which performs computation,
then when both of the above assumptions are translated into the language
of physics, they embody what is known as the ``Bekenstein
bound''~\cite{PhysRevD.23.287}, which is an upper limit on the amount
of information that can be contained within a given finite region of
space. A detailed historical account of these ideas in the context of
modern physics is again beyond our scope. However, the quotes above,
like the ones before, should convey the ideas that our theories of
computation and complexity are based on some physical assumptions that
Turing and others found reasonable in the 1930s.

To summarize, a major achievement of computer science has been the
development of abstract models of computation that shield the
discipline from rapid changes in the underlying technology. Yet, as
effective as these models have been, one must note that they
\emph{embody several implicit physical assumptions} and these
assumptions are based on a certain understanding of the laws of
physics. Our understanding of physics has evolved tremendously since
1900!  Thus it is time to revisit these abstractions, especially with
respect to quantum mechanics.  Indeed one should take the physical
principles underlying quantum mechanics, the most successful physical
theory known to us, and adapt computation to ``learn'' from these
principles.  In the words of Girard~\cite{Girard:2007:TMI:1348911.1348915}:
\begin{quote}
  In other terms, what is so good in logic that quantum physics should
  obey?  Can't we imagine that our conceptions about logic are wrong,
  so wrong that they are unable to cope with the quantum miracle?
  [\ldots] Instead of teaching logic to nature, it is more reasonable
  to learn from her. Instead of interpreting quantum into logic, we
  shall interpret logic into quantum.
\end{quote}

There are, in fact, many different quantum mechanical principles which
are at odds with our current models of computation. In this paper, we
will focus on the previously identified principle of
\emph{reversibility}. In more detail, we will view data as an explicit
representation of \emph{information} and programs as processes that
transform information in a reversible way, i.e., processes that are
subject to the physical principle of \emph{conservation of
  information.} We will formalize this idea and follow its
consequences, which will turn out to be far reaching.

\paragraph*{Programs as Reversible Deformations.}
To better understand the essence of ``conservation of information'' in
the context of computing, we first look for analogous ideas in
physics, but this time at the macro scale. Viewing information as a
physical object, what does it mean to transform an object in such a
way that we do not lose its fundamental character?

For rigid objects (like a chair), the only such transformations are
translations and rotations. But what about something more flexible,
with multiple representations, such as a water balloon?  Such objects
can be \emph{deformed} in various ways, but still retain their
fundamental character -- as long as we do not puncture them or
over-stretch them. Ignoring material characteristics
(i.e. over-stretching), what is special about these deformations, as
well as for translations and rotations, is that they correspond to
continuous maps, with a continuous inverse. In fact, even more is
true: they are analytic maps, with analytic inverses. For our purpose,
the most important part is that such maps are infinitely
differentiable.  In other words, not only is there an inverse to the
deformation, but its derivative is also invertible, and so on.

When we look around, we find many different words for related
concepts: isomorphism, equivalence, sameness, equality,
interchangeability, comparability, and correspondence, to name a
few. Some of these are informal concepts, while others have formal
mathematical meaning.  More importantly, even amongst the formal
concepts, there are differences -- which is why there are so many of
them! Because there are many such notions, we also need to walk our
way through them to find the one which is ``just right.'' Thus we seek
a concept which is neither too strong nor too weak, that will express
when some structured information should be treated as ``the same.''
We can draw an analogy with topology: in topology, all point sets can
always be equipped with either the discrete or the indiscrete
topology, but both of these extremes are rarely useful. We will
develop our working notion of ``sameness'' as we go through the
various components that make up a programming language.

Starting from the physical perspective, whatever our notion of data
is, we will be interested in programs as representing transformations
of that data which are reversible. In other words, we want our
programs-as-transformations to ``play well'' with the inherent notion
of ``sameness'' that our data will carry. Thus we need to start by
looking at what structure our data has, which will help us define
an appropriate notion of a reversible program. Of course, when
programs themselves are data, things do get more complicated.  In the
following sections, we will look at different natural classes of data,
and explore the corresponding notion of reversible programs.

To summarize, we will take ``the same'' as a fundamental principle and
derive what it means for data, programs, program transformations, as
well as proofs / deductions, to be ``the same'' -- in a manner
consistent with preservation of information. This stands in stark
contrast with most current approaches to reversible computation, which
start from current models of computation involving irreversible
operations and try to find various ways to \emph{patch things up} so
as to be reversible.

\paragraph*{Reversible Programming Languages.} The practice of
programming languages is replete with \emph{ad hoc} instances of
reversible computations: database transactions, mechanisms for data
provenance, checkpoints, stack and exception traces, logs, backups,
rollback recoveries, version control systems, reverse engineering,
software transactional memories, continuations, backtracking search,
and multiple-level undo features in commercial applications. In
the early nineties, Baker~\cite{Baker:1992:LLL,Baker:1992:NFT} argued
for a systematic, first-class, treatment of reversibility. But
intensive research in full-fledged reversible models of computations
and reversible programming languages was only sparked by the discovery
of deep connections between physics and
computation~\cite{Landauer:1961,PhysRevA.32.3266,Toffoli:1980,bennett1985fundamental,Frank:1999:REC:930275},
and by the potential for efficient quantum
computation~\cite{springerlink:10.1007/BF02650179}.

The early developments of reversible programming languages started
with a conventional programming language, e.g., an extended
$\lambda$-calculus, and either
\begin{enumerate}
\item extended the language with a history
mechanism~\cite{vanTonder:2004,Kluge:1999:SEMCD,lorenz,danos2004reversible}, or
\item imposed constraints on the control flow constructs to make them
reversible~\cite{Yokoyama:2007:RPL:1244381.1244404}.
\end{enumerate}
More modern approaches recognize that reversible programming languages require
a fresh approach and should be designed from first principles without the
detour via conventional irreversible
languages~\cite{Yokoyama:2008:PRP,Mu:2004:ILRC,abramsky2005structural,DiPierro:2006:RCL:1166042.1166047}.

In previous work, Carette, Bowman, James, and
Sabry~\cite{rc2011,James:2012:IE:2103656.2103667,Carette2016}
introduced the~$\Pi$ family of typed reversible languages.  As motivated above,
the starting point for this development is the physical principle of
\emph{conservation of
information}~\cite{Hey:1999:FCE:304763,fredkin1982conservative} and
the family of languages is designed to embrace this principle by
requiring all computations to preserve information.

The fragment without
recursive types is universal for reversible boolean
circuits~\cite{James:2012:IE:2103656.2103667} and the extension with
recursive types and trace
operators~\cite{Hasegawa:1997:RCS:645893.671607} is a Turing-complete
reversible language~\cite{James:2012:IE:2103656.2103667,rc2011}. While
at first sight, $\Pi$ too might appear \emph{ad hoc}, it really arises
naturally from an ``extended'' view of the Curry-Howard
correspondence~\cite{Carette2016}: rather than looking at mere
\emph{inhabitation} as the main source of analogy between logic and
computation, \emph{type equivalence} becomes the source of analogy.
Taking inspiration from the fact that many terms of the $\lambda$-calculus
arise from Cartesian Closed Categories including, most importantly,
a variety of propositional equalities and computation rules,
allows us to pursue that analogy further. Some of the details
of this development will be motivated and explained in the
present paper.

\section{Data I: Finite Sets}
\label{sec:dataone}

Most programming languages provide primitive data like booleans,
characters, strings, and (bounded) numbers that are naturally modeled
as finite sets. We therefore start by modeling reversible computations
over finite and discrete spaces of points. Infinite sets are more
subtle, and will be discussed in the conclusion.

What does it mean to deform a space of points? For example, what
transformation can we do on a bag of marbles? Well, we can shuffle
them around and that is the only transformation that will preserve the
space. Turning to the mathematical abstraction as sets, we ask what
does it mean for two finite sets to be ``the same''?  Well, clearly
the sets $A = \left\{1, 2, 3\right\}$ and $B = \left\{c, d\right\}$
are different.  Why?  Well, suppose there was a transformation
$f : A \rightarrow B$ that deformed $A$ into $B$, and another
$g : B \rightarrow A$ which undid this transformation. Since $f$ is
total, by the pigeonhole principle, two elements of $A$ would be
mapped to the same element of $B$. Suppose that this is $2$ and $3$,
and that they both map to $d$.  But $g(d)$ cannot be both $2$ and $3$,
and so $g$ is not the inverse of $f$. With just a little more work, we
can show that $f$ (and $g$) must be both injective and surjective. In
other words, $f$ (and $g$) must be a bijection between $A$ and $B$.
And of course this only happens when $A$ and $B$ have the same number
of elements. More importantly, given a bijection $f : C \rightarrow D$
of finite sets $C,D$, there always exists another bijection
$g : D \rightarrow C$ which is $f$'s inverse. So, for finite sets,
\emph{bijections} act as reversible deformations.

This discussion is purely ``semantic,'' in the sense that it is about
the denotation of simple primitive data (sets) and their reversible
deformations (bijections).  We would like to reverse engineer a programming
language from this denotation. But first, an obvious remark: any two sets
$C$ and $D$ of cardinality $n$ are always in bijective correspondence. So
we can abstract away from the details of the elements of $C$ and $D$ and
instead choose canonical representations -- in much the same way as computers
choose binary words to represent everything.

\begin{defn} For $n\in\Nat$, denote by $\fin{n}$ the set
$\left\{0,1,\ldots,n-1\right\}$.
We will refer to $\fin{n}$ as the canonical set with $n$ elements.
\end{defn}

Bijections on \fin{n} have a specific name: permutations. As is
well-known, permutations can be generated by sequential compositions of
transpositions. Thus we can create a small language for writing
permutations on $\fin{n}$ as:
\[\begin{array}{rcl}
p^n &::=& \mathit{id} ~\mid~ \mathit{swap}~i~j ~\mid~ p^n \fatsemi p^n
  \end{array}\]
where $i,j:\Nat$, $i\neq j$ and $i,j < n$. Note that we could remove
$\mathit{id}$ from the language and drop the $i\neq j$ condition so that
$\mathit{swap}~j~j$ would represent the identity permutation.

For convenience, we write $[2^n]$ for the finite set representing
$n$-bit words with the canonical ordering for binary numbers. Thus
when $n=3$, the finite set has elements $\{0,1,2,3,4,5,6,7\}$ which
correspond to the 3-bit words $\{000,001,010,011,100,101,110,111\}$.
Although this language appears weak, it is universal for reversible
boolean combinational circuits $[2^i] \rightarrow [2^i]$ with $i$
input/output wires.

To illustrate the expressiveness of the language, we develop a few
small examples. We start by writing boolean negation ``not'' as a
permutation $[2^1] \rightarrow [2^1]$, the controlled-not gate (also
known as ``cnot'') as a permutation $[2^2] \rightarrow [2^2]$, and the
controlled-controlled-not gate (also known as ``toffoli'') as a
permutation $[2^3] \rightarrow [2^3]$:
\[\begin{array}{rcl}
\mbox{not} &=& \mathit{swap}~0~1 \\
\mbox{cnot} &=& \mathit{swap}~2~3 \\
\mbox{toffoli} &=& \mathit{swap}~6~7
\end{array}\]
The ``cnot'' gate operates on two bits and negates the second (the
target bit) if the first one (the control bit) is 1, i.e., it swaps
$10$ and $11$; the ``toffoli'' gate negates the third bit (the target
bit) if both the first two bits (the control bits) are 1, i.e., it
swaps $110$ and $111$.

There is however a subtle issue: programming in such an unstructured
language is \emph{not} compositional in the sense that using the
``not'' gate in a larger circuit forces us to change its
implementation. Indeed if we had two bits and wanted to use ``not'' to
negate the first bit, we would write the permutation of type
$[2^2] \rightarrow [2^2]$ that permutes $00$ with $10$ \emph{and}
permutes $01$ with $11$, i.e, the permutation
$\mathit{swap}~0~2 \fatsemi \mathit{swap}~1~3$. To illustrate
how inconvenient this is, consider the reversible full adder
below designed by Desoete et al.~\cite{117414}:

\begin{center}
\includegraphics[scale=0.6]{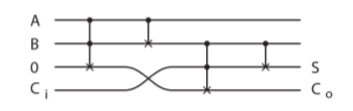}
\end{center}

In the figure (copied from a more general paper that includes
alternative designs~\cite{Rentergem2005OptimalDO}), the full adder
takes 4 inputs: the two bits to add $A$ and $B$, an incoming carry bit
$C_i$, and a heap input initialized to 0 to maintain
reversibility. There are four outputs: the first two are identical to
the incoming bits $A$ and $B$ and are considered ``garbage.'' The
third bit $S$ is the sum and the last bit $C_o$ is the outgoing carry
bit. In the notation used to describe the circuit, the $\times$ denotes
boolean negation and the dots are control bits. In our reversible
language, we can express this circuit as the following permutation of
type $[2^4] \rightarrow [2^4]$:

\begin{equation}\label{eq:adder}
\begin{array}{l@{\hspace{1cm}}l}
\mathit{swap}~12~14 \fatsemi \mathit{swap}~13~15 ~\fatsemi & \mbox{toffoli} \\
\mathit{swap}~8~12 \fatsemi \mathit{swap}~9~14 \fatsemi
    \mathit{swap}~10~13 \fatsemi \mathit{swap}~11~15 ~\fatsemi & \mbox{cnot and swap} \\
\mathit{swap}~6~7 \fatsemi \mathit{swap}~14~15 ~\fatsemi & \mbox{toffoli} \\
\mathit{swap}~4~6 \fatsemi \mathit{swap}~5~7 \fatsemi
    \mathit{swap}~12~14 \fatsemi \mathit{swap}~13~15 & \mbox{cnot}
\end{array}
\end{equation}

Note how the implementation of $\mathit{cnot}$ as a permutation
$[2^2] \rightarrow [2^2]$ cannot be directly reused in the larger
circuit $[2^4] \rightarrow [2^4]$.

For such reasons, in programming practice we are interested in
structured data and compositional abstractions, which will be the
subject of the next section. What we do learn from this short investigation
using untyped and unstructured sets is what the ``\emph{purely
  operational}'' view of the theory would be. In particular, it tells
us that permutations are an inescapable part of the fabric of
reversible computing.  However as permutations are untyped, and act on
the canonicalized version of $n$-element sets (i.e. those sets where
all the structure has been forgotten), these are a rather pale shadow
of the rich tapestry of information-preserving transformations of
structured data, which we investigate next.

\section{Data II: Structured Finite Types}
\label{sec:pi1}

Instead of spaces (aka discrete sets) consisting solely of
unstructured isolated points, we now investigate structured spaces
built from sums and products of elementary spaces. This structure
corresponds to the building blocks of type theory which are: the empty
type ($\bot$), the unit type ($\top$), the sum type ($\presumtype$), and
the product ($\preprodtype$) type. Before getting into the formal theory,
let's consider possible deformations on the space $(\top \presumtype \bot)
\preprodtype (\top \presumtype \top)$. This space is the product of two
subspaces: the subspace $(\top \presumtype \bot)$ which itself is the sum
of the space~$\top$ containing one element $\texttt{tt}$ and the empty
space $\bot$ and the subspace $(\top \presumtype \top)$ which is the sum of
two spaces each containing the one element $\texttt{tt}$. First, as
discussed in the previous section, any deformation of this space must
at least preserve the number of elements: we can neither create nor
destroy points during any continuous deformation. Seeing that the
number of elements in our example space is 2, a reasonable hypothesis
is that we can deform the space above to any other space with 2
elements such as $\top \presumtype \top$ or $\top \presumtype (\top \presumtype
\bot)$. What this really means is that we are treating the sum and
product structure as malleable. For example, imagining a product
structure as arranged in a grid; by ``stretching'' we can turn it
in to a sum structure arranged in a line. We can also change the
orientation of the grid by exchanging the axes, as well as do other
transformations --- as long as we preserve the number of points.
Of course, it is not a priori clear that this necessary requirement
is also sufficient.  Making this intuition precise will be the topic of this
section.

\subsection{A Model of Type Equivalences}

We now want a proper mathematical description of this idea. Our goal
is a denotational semantics on types which makes types that have the
same number of points be equivalent types.  First we note that the
structure of types has a nice correspondence (Curry-Howard) to logic:

\begin{center}
\begin{tabular}{c|c}
Logic & Types \\ \hline
$\false$ & $\bot$ \\
$\true$ & $\top$ \\
$\land$ & $\preprodtype$ \\
$\lor$ & $\presumtype$ \\
\end{tabular}
\end{center}

\noindent This correspondence is rather fruitful. As logical
expressions form a commutative semiring, we would expect that types
too form a commutative semiring. And indeed they do -- at least up to
\emph{type isomorphism}.  The natural numbers $\Nat$ are another
commutative semiring; it will turn out that, even though the
Curry-Howard correspondence has been extremely fruitful for
programming language research, it is $\Nat$ which will be a better
model for finite structured types as the corresponding commutative
semiring captures the familiar numerical identities that preserve the
number of points in the types.

\begin{defn}
\label{def:rig}
A \emph{commutative semiring} (sometimes called a \emph{commutative
  rig} --- commutative ri\emph{n}g without negative elements)
$(R,0,1,+,\cdot)$ consists of a set $R$, two distinguished elements of
$R$ named $0$ and $1$, and two binary operations~$+$ and $\cdot$,
satisfying the following relations for any $a,b,c \in R$:
\[\begin{array}{rcl}
0 + a &=& a \\
a + b &=& b + a \\
a + (b + c) &=& (a + b) + c \\
\\
1 \cdot a &=& a \\
a \cdot b &=& b \cdot a \\
a \cdot (b \cdot c) &=& (a \cdot b) \cdot c \\
\\
0 \cdot a &=& 0 \\
(a + b) \cdot c &=& (a \cdot c) + (b \cdot c)
\end{array}\]
\end{defn}

\begin{prop}
The structure $\left(\{\false,\true\}, \false, \true, \lor, \land\right)$
is a commutative semiring.
\end{prop}

We would like to adapt the commutative semiring definition to the
setting of structured types. First, types do not naturally want to be
put together into a ``set.''  This can be fixed if we replace the set
$R$ with a universe $U$, and replace the set membership $0 \in R$ with
the typing judgement $\bot : U$ (and similarly for the other
items). Our next instinct would be to similarly replace $=$ with a
type $A \equiv B$ that asserts that $A$ and $B$ are
\emph{propositionally equal}, i.e. reduce to equivalent type-denoting
expressions under the rules of the host type system.  This is however
not true: the proposition $A \preprodtype B \equiv B \preprodtype A$ is not
normally\footnote{Except in univalent type theory where equivalent
  types are identified.} provable for arbitrary types $A$ and $B$. But
it should be clear that $A \preprodtype B$ and $B \preprodtype A$ contain
equivalent information. In other words, we would like to be able to
witness that $A \preprodtype B$ can be reversibly deformed into
$B \preprodtype A$, and vice-versa, which motivates the introduction of type
\emph{equivalences}. To do this, we need a few important auxiliary
concepts.

\begin{defn}[Propositional Equivalence]\label{def:propeq}
Two expressions $a, b$ of type $A$ are \emph{propositionally
equal} if their normal forms are equivalent under the rules
of the host type system.
\end{defn}

In Martin-L\"{o}f Type Theory, normal forms mean $\beta\eta$-long
normal forms under $\alpha$-equivalence. In other words, expressions
are evaluated as much as possible ($\beta$-reduced), all functions are
fully applied ($\eta$-long), and the exact names of bound variables
are irrelevant ($\alpha$-equivalence). Note that the above definition
applies equally well to expressions that denote values and expressions
that denote types.

\begin{defn}[Homotopy]
\label{def:homotopy}
Two functions $f,g:A \rightarrow B$ are \emph{homotopic}
if~$\forall x:A. f(x) \equiv g(x)$. We denote this $f \sim g$.
\end{defn}

\noindent It is easy to prove that homotopies (for any given function
space $A \rightarrow B$) are an equivalence relation.  The simplest
definition of the data which makes up an equivalence is the following.

\begin{defn}[Quasi-inverse]
\label{def:quasi}
For a function $f : A \rightarrow B$, a \emph{quasi-inverse} is a
triple $(g, \alpha, \beta)$, consisting of a function
$g : B \rightarrow A$ and two homotopies
$\alpha : f \circ g \sim \mathrm{id}_B$ and
$\beta : g \circ f \sim \mathrm{id}_A$.
\end{defn}

\begin{defn}[Equivalence of types]
  Two types $A$ and $B$ are equivalent $A \simeq B$ if there exists a
  function $f : A \rightarrow B$ together with a quasi-inverse for $f$.
\end{defn}

\noindent Why \emph{quasi}? The reasons are beyond our scope, but the
interested reader can read Sec.~$2.4$ and Ch.~$4$ in the
Homotopy Type Theory (HoTT) book~\cite{hottbook}.
There are several conceptually different, but
equivalent, ``better'' definitions.  We record just one here:

\begin{defn}[Bi-invertibility]
\label{def:biinv}
For a function $f : A \rightarrow B$, a \emph{bi-inverse} is a
pair of functions $g,h : B \rightarrow A$ and two homotopies
$\alpha : f \circ g \sim \mathrm{id}_B$ and
$\beta : h \circ f \sim \mathrm{id}_A$.
\end{defn}


\noindent We can then replace quasi-inverse with bi-invertibility in
the definition of type equivalence. The differences will not matter to
us here.


We are now in position to describe the commutative
semiring structure for types. After replacing the set $R$ with a
universe $U$, we also replace the algebraic use of $=$ in
Def.~\ref{def:rig} by the type equivalence relation $\simeq$. With
this change, we can indeed prove that types (with $\bot, \top, \presumtype,
\preprodtype$) form a commutative semiring. The reader familiar with
universal algebra should pause and ponder a bit about what we have
done. We have lifted \emph{equality} from being in the signature of
the ambient logic and instead put it in the signature of the algebraic
structure of interest.  In simpler terms, we shift equality from
having a privileged status in our meta-theory, to being just another
symbol (denoting an equivalence relation) in our theory.  The understanding
that equality is not an absolute concept has recently been an area of
active research in mechanized mathematics --- although the concepts of
intensional versus extensional equality go back to Frege and Russell.

If we revisit the Curry-Howard correspondence, we notice one
more issue. In logic, it is true that $A \lor A = A$ and
$A \land A = A$. However, neither $A \presumtype A$ nor $A \preprodtype A$ are
equivalent to $A$. They are however \emph{equi-inhabited}. This is
a fancy way of saying
\[ A \presumtype A \ \text{is inhabited} \qquad \Leftrightarrow \qquad A \
  \text{is inhabited} \] The above is the real \textit{essence} of the
Curry-Howard correspondence.  In other words, classical Curry-Howard
tells us about \emph{logical equivalence} of types. This is even a
constructive statement: there are indeed functions
$f : A \presumtype A \rightarrow A$ and $g : A \rightarrow A \presumtype A$;
however, they are not inverses.

So mere inhabitation falls far short of our goals of being able to
smoothly deform from one type to another. Let us thus analyze the crux
of the ``problem.'' In logic, we have that $\land$ and $\lor$ are both
\emph{idempotent}: this is the property of any binary operation $\circ$
where $\forall a. a \circ a = a$. And it should be clear that an
idempotent operations is a \emph{forgetful} operation: its input has
two copies of $a$, but its output, only one. On the type side,
something more subtle happens. Consider $\top \presumtype \top$ versus
$\top$; the first has exactly \emph{two} proofs of inhabitation
($\inleft{\texttt{tt}}$ and $\inright{\texttt{tt}}$) while the second
only one ($\texttt{tt}$). These cannot be put in bijective
correspondence. Even though the ``payload'' \texttt{tt} is the same,
forgetting $\texttt{left}$ (or \texttt{right}) throws away information
-- something we have expressly disallowed.  Yes, this should remind
you of Maxwell's daemon: even though the data is the same, they are
tagged differently, and these tags are indeed information, and their
information content must be preserved.

\begin{figure}[t]
\[
\begin{array}{rrcll}
& A & \simeq & A &\\
\\
&  \bot \presumtype A & \simeq & A &\\
&  A \presumtype B & \simeq & B \presumtype A &\\
&  A \presumtype (B \presumtype C) & \simeq & (A \presumtype B) \presumtype C &\\
\\
&  \top \preprodtype A & \simeq & A &\\
&  A \preprodtype B & \simeq & B \preprodtype A &\\
&  A \preprodtype (B \preprodtype C) & \simeq & (A \preprodtype B) \preprodtype C &\\
\\
& \bot \preprodtype A & \simeq & \bot &\\
& (A \presumtype B) \preprodtype C & \simeq & (A \preprodtype C) \presumtype (B \preprodtype C) &
\end{array}
\]
\caption{Type isomorphisms.}
\label{type-isos}
\end{figure}


Nevertheless, the Curry-Howard correspondence still has some force. We
know that the inhabitants of types formed with with $\bot, \top,
\presumtype, \preprodtype$ form a commutative semiring. What we want
to know is, which types are equivalent? From a commutative semiring
perspective, this amounts to asking what terms are equal.  We have a
set of generators for those equations, namely those in
Def.~\ref{def:rig}. What we thus need is to create $8$ pairs of
mutually inverse functions which witness these identities.  For
concreteness, we show the signatures in Fig.~\ref{type-isos}.

From category theory, we are informed of the following privilege
enjoyed by the natural numbers~$\Nat$:
\begin{thm}
  The semiring $\left(\Nat, 0, 1, +, \cdot\right)$ is \emph{initial}
  in the category of semirings and semiring homomorphisms.
\end{thm}
\noindent In other words, for any semiring $S$, there is a homomorphism
from $\Nat$ into $S$. But $\Nat$ is also the ``counting'' semiring,
which formalizes the notion of cardinality of finite discrete sets.

The previous section on finite sets, along with the reasoning above,
thus leads us to posit that the correct denotational semantics for
finite discrete types is that of the semiring $\left(\Nat, 0, 1, +,
\cdot\right)$. It is worth noting that equality in this semiring is
intensional (i.e. two things are equal if and only if they are
identical after evaluation), unlike that for types.

\subsection{A Language of Type Equivalences}

\begin{figure}[t]
\[
\begin{array}{rrcll}
\idc :& t & \iso & t &: \idc \\
\\
\identlp :&  0 \sumtype t & \iso & t &: \identrp \\
\swapp :&  t_1 \sumtype t_2 & \iso & t_2 \sumtype t_1 &: \swapp \\
\assoclp :&  t_1 \sumtype (t_2 \sumtype t_3) & \iso & (t_1 \sumtype t_2) \sumtype t_3 &: \assocrp \\
\\
\identlt :&  1 {\prodtype} t & \iso & t &: \identrt \\
\swapt :&  t_1 {\prodtype} t_2 & \iso & t_2 {\prodtype} t_1 &: \swapt \\
\assoclt :&  t_1 {\prodtype} (t_2 {\prodtype} t_3) & \iso & (t_1 {\prodtype} t_2) {\prodtype} t_3 &: \assocrt \\
\\
\absorbr :&~ 0 {\prodtype} t & \iso & 0 ~ &: \factorzl \\
\dist :&~ (t_1 \sumtype t_2) {\prodtype} t_3 & \iso & (t_1 {\prodtype} t_3) \sumtype (t_2 {\prodtype} t_3)~ &: \factor
\end{array}
\]
\caption{$\Pi$-terms.}
\label{pi-terms}
\end{figure}

We now have in our hands our desired denotational semantics for types.
We want to create a programming language, which we call $\Pi$, such
that the types and type combinators map to $\bot, \top, \presumtype,
\preprodtype$, and such that we have ground terms whose denotation are
all $16$ type isomorphisms of Fig.~\ref{type-isos}. This is rather
straightforward, as we can simply do this literally. To make the
analogy with commutative semirings stand out even more, we will use
$0, 1, \sumtype$, and ${\prodtype}$ at the type level, and will denote
``equivalence'' by $\iso$.  Thus Fig.~\ref{pi-terms} shows the
``constants'' of the language.  As these all come in symmetric pairs
(some of which are self-symmetric), we give names for both directions.
Note how we have continued with the spirit of Curry-Howard: the terms
of $\Pi$ are \emph{proof terms}, but rather than being witnesses of
inhabitation, they are witnesses of equivalences. Thus we get an
unexpected programming language design:

\begin{center}
\fbox{ The proof terms denoting commutative semiring equivalences
  induce the terms of $\Pi$.}
\end{center}
\vspace*{3mm}

\begin{figure}[t]
\[
\Rule{}
{\jdg{}{}{c_1 : t_1 \iso t_2} \quad \vdash c_2 : t_2 \iso t_3}
{\jdg{}{}{c_1 \odot c_2 : t_1 \iso t_3}}
{}
\qquad
\Rule{}
{\jdg{}{}{c_1 : t_1 \iso t_2} \quad \vdash c_2 : t_3 \iso t_4}
{\jdg{}{}{c_1 \oplus c_2 : t_1 \sumtype t_3 \iso t_2 \sumtype t_4}}
{}
\]
\[
\Rule{}
{\jdg{}{}{c_1 : t_1 \iso t_2} \quad \vdash c_2 : t_3 \iso t_4}
{\jdg{}{}{c_1 \otimes c_2 : t_1 {\prodtype} t_3 \iso t_2 {\prodtype} t_4}}
{}
\]
\caption{$\Pi$-combinators.}
\label{pi-combinators}
\end{figure}

\noindent
Of course, one does not get a programming language with just typed
constants! There is a need to perform multiple equivalences. There are
in fact three ways to do this: sequential composition $\odot$, choice
composition $\oplus$ (sometimes called juxtaposition), and parallel
composition $\otimes$. See Fig.~\ref{pi-combinators} for the
signatures. The construction $c_1 \odot c_2$ corresponds to performing
$c_1$ first, then $c_2$, and is the usual notion of composition -- and
corresponds to $\fatsemi$ of the language of permutations of
Sec.~\ref{sec:dataone}. The construction $c_1 \oplus c_2$ chooses to
perform $c_1$ or $c_2$ depending on whether the input is labelled
$\textsf{left}$ or $\textsf{right}$ respectively. Finally the
construction $c_1 \otimes c_2$ operates on a product structure, and
applies $c_1$ to the first component and $c_2$ to the second. The
language of permutations lacked the ability to combine permutations by
taking sums and products, which led to the awkward non-compositional
programming style illustrated in the full adder
example~(Eq.~\ref{eq:adder}).


Thus the denotation of the $\Pi$ terms \emph{should} be
permutations. But given types $A$ and $B$ denoting $\fin{m}$
and~$\fin{n}$ respectively, what are $A \presumtype B$ and $A \preprodtype B$ ?
They correspond exactly to $\fin{m+n}$ and $\fin{m*n}$!
Geometrically, this corresponds to concatenation for $A + B$,
i.e. lining up the elements of $A$ first, and then those of~$B$. For
$A * B$, one can picture this as lining up the elements of $A$
horizontally, those of $B$ vertically and perpendicular to those of
$A$, and filling in the square with pairs of elements from $A$ and
$B$; if one re-numbers these sequentially, reading row-wise, this
gives an enumeration of $\fin{m*n}$.

From here, it is easy to see what, for example, $c_1 \oplus c_2$ must be,
operationally: from a permutation on $\fin{m}$ and another on $\fin{n}$,
create a permutation on $\fin{m+n}$ by having $c_1$ operate on the first
$m$ elements of $A+B$, and $c_2$ operate on the last $n$ elements.
Similarly, $\swapp$ switches the roles of $A$ and $B$, and thus corresponds
to $\fin{n+m}$. Note how we ``recover'' the commutativity of
natural number addition from this type isomorphism. Geometrically, $\swapt$
is also rather interesting: it corresponds to matrix transpose!
Furthermore, in this representations, some combinators like
$\identlp$ and $\assoclp$ are identity operations: the underlying representations
are not merely isomorphic, they are definitionally equal.
In other words, the passage to $\Nat$ erases some structural information.

\begin{figure}[t]
\[
\Rule{}
{\jdg{}{}{c_1 : t_1 \iso t_2}}
{\jdg{}{}{\ !\ c_1 : t_2 \iso t_1}}
{}
\]
\caption{Derived $\Pi$-combinator.}
\label{derived-pi-combinator}
\end{figure}

Embedded in our definition of $\Pi$ is a conscious design decision: to make the
terms of $\Pi$ \emph{syntactically} reversible. In other words, to
every $\Pi$ constant, there is another $\Pi$ constant which is its
inverse. As this is used frequently, we give it the short name $!$,
and its type is given in Fig.~\ref{derived-pi-combinator}. This
combinator is \emph{defined}, by pattern matching on the syntax of
its argument and structural recursion.

This is not the only choice.  Another would be to add a
$\mathit{flip}$ combinator to the language; we could then remove
quite a few combinators as redundant. The drawback is that many
programs in $\Pi$ become longer. Furthermore, some of the symmetry
at ``higher levels'' (see next section) is also lost. Since the
extra burden of language definition and of proofs is quite low, we
prefer the structural symmetry over a minimalistic language definition.

\begin{figure}[t]
\[
\begin{array}{rrcll}
\identlsp :&  t \sumtype 0 & \iso & t &: \identrsp \\
\identlst :&  t {\prodtype} 1 & \iso & t &: \identrst \\
\\
\absorbl :&~ t {\prodtype} 0 & \iso & 0 ~ &: \factorzr \\
\distl :&~ t_1 {\prodtype} (t_2 \sumtype t_3) & \iso & (t_1 {\prodtype} t_2) \sumtype (t_1 {\prodtype} t_3)~ &: \factorl
\end{array}
\]
\caption{Additional $\Pi$-terms.}
\label{more-pi}
\end{figure}

We also make a second design decision, which is to make the $\Pi$
language itself symmetric in another sense: we want both left
and right introduction/elimination rules for units, $0$ absorption
and distributivity. Specifically, we add the $\Pi$-terms of
Fig.~\ref{more-pi} to our language. These are redundant because
of $\swapp$ and $\swapt$, but will later enable shorter programs
and more elegant presentation of program transformations.

This set of isomorphisms is known to be sound and
complete~\cite{Fiore:2004,fiore-remarks} for isomorphisms
of finite types.  Furthermore, it is also universal
for hardware combinational
circuits~\cite{James:2012:IE:2103656.2103667}.

\subsection{Operational Semantics}
\label{sec:opsem}

To give an operational semantics to $\Pi$, we are mainly missing
a notation for \emph{values}.

\begin{definition}{(Syntax of values of \ensuremath{\Pi })}
\label{def:langRev}
\[\begin{array}{rclr}
 \mathit{values}, v &::=& () ~|~ \mathit{left} ~v ~|~ \mathit{right} ~v ~|~ (v,v) \\
 \end{array}\]
\end{definition}

Given a program \ensuremath{c : b_1 \leftrightarrow b_2} in \ensuremath{\Pi },
we can run it by supplying it with a value \ensuremath{ v_1 : b_1}. The
evaluation rules \ensuremath{c ~v_1 \mapsto v_2} are given below.

\begin{definition}{(Operational Semantics for \ensuremath{\Pi })}
\label{def:operational-langRev}

Identity:
\[\begin{array}{rlcl}
 \idc & v &\mapsto & v \\
 \end{array}\]

Additive fragment:
\[\begin{array}{rlcl}
 \identlp & (\mathit{right} ~v) &\mapsto & v \\
 \identrp & v &\mapsto & \mathit{right} ~v \\
 \identlsp & (\mathit{left} ~v) &\mapsto & v \\
 \identrsp & v &\mapsto & \mathit{left} ~v \\
 \swapp & (\mathit{left} ~v) &\mapsto & \mathit{right} ~v \\
 \swapp & (\mathit{right} ~v) &\mapsto & \mathit{left} ~v \\
 \assoclp & (\mathit{left} ~v_1) &\mapsto & \mathit{left} ~(\mathit{left} ~v_1) \\
 \assoclp & (\mathit{right} ~(\mathit{left} ~v_2)) &\mapsto & \mathit{left} ~(\mathit{right} ~v_2) \\
 \assoclp & (\mathit{right} ~(\mathit{right} ~v_3)) &\mapsto & \mathit{right} ~v_3 \\
 \assocrp & (\mathit{left} ~(\mathit{left} ~v_1)) &\mapsto & \mathit{left} ~v_1 \\
 \assocrp & (\mathit{left} ~(\mathit{right} ~v_2)) &\mapsto & \mathit{right} ~(\mathit{left} ~v_2) \\
 \assocrp & (\mathit{right} ~v_3) &\mapsto & \mathit{right} ~(\mathit{right} ~v_3) \\
 \end{array}\]

Multiplicative fragment:
\[\begin{array}{rlcl}
 \identlt & ((), v) &\mapsto & v \\
 \identrt & v &\mapsto & ((), v) \\
 \identlst & (v, ()) &\mapsto & v \\
 \identrst & v &\mapsto & (v, ()) \\
 \swapt & (v_1, v_2) &\mapsto & (v_2, v_1) \\
 \assoclt & (v_1, (v_2, v_3)) &\mapsto & ((v_1, v_2), v_3) \\
 \assocrt & ((v_1, v_2), v_3) &\mapsto & (v_1, (v_2, v_3)) \\
 \absorbr & (v_1, v_2) & \mapsto & v_1 \\
 \end{array}\]

Distributivity and factoring:

\[\begin{array}{rlcl}
 \dist & (\mathit{left} ~v_1, v_3) &\mapsto & \mathit{left} ~(v_1, v_3) \\
 \dist & (\mathit{right} ~v_2, v_3) &\mapsto & \mathit{right} ~(v_2, v_3) \\
 \distl & (v_1, \mathit{left} ~v_2) &\mapsto & \mathit{left} ~(v_1, v_2) \\
 \distl & (v_1, \mathit{right} ~v_3) &\mapsto & \mathit{right} ~(v_1, v_3) \\
 \factor & (\mathit{left} ~(v_1, v_3)) &\mapsto & (\mathit{left} ~v_1, v_3) \\
 \factor & (\mathit{right} ~(v_2, v_3)) &\mapsto & (\mathit{right} ~v_2, v_3) \\
 \factorl & (\mathit{left} ~(v_1, v_2)) &\mapsto & (v_1, \mathit{left} ~v_2) \\
 \factorl & (\mathit{right} ~(v_1, v_3)) &\mapsto & (v_1, \mathit{right} ~v_3) \\
 \absorbl & (v_1 , v_2) & \mapsto & v_2 \\
 \end{array}\]

The evaluation rules of the composition combinators are given below:

$$
\infer{ (c_1\odot c_2) ~v_1 \mapsto v_2}{
	 c_1 ~v_1 \mapsto v
	&
	 c_2 ~v \mapsto v_2
}
$$
$$
\infer{ (c_1 \oplus c_2) ~(\mathit{left} ~v_1) \mapsto \mathit{left} ~v_2}{
	 c_1 ~v_1 \mapsto v_2
}
\quad
\infer{ (c_1 \oplus c_2) ~(\mathit{right} ~v_1) \mapsto \mathit{right} ~v_2}{
	 c_2 ~v_1 \mapsto v_2
}
$$
$$
\infer{ (c_1 \otimes c_2) ~(v_1, v_2) \mapsto (v_3, v_4)}{
	 c_1 ~v_1 \mapsto v_3
	&
	 c_2 ~v_2 \mapsto v_4
}
$$

\end{definition}

Since there are no values that have the type \ensuremath{0}, the
reductions for the combinators \identlp, \identrp, \identlsp, and
\identrsp\ omit the impossible cases. \factorzr\ and \factorzl\
likewise do not appear as they have no possible cases at all. However,
\absorbr\ and \absorbl\ are treated slightly differently: rather than
\emph{eagerly} assuming they are impossible, the purported inhabitant
of $0$ given on one side is passed on to the other side. The reason
for this choice will have to wait for Sec.~\ref{langeqeq} when we
explain some higher-level symmetries (see Fig.~\ref{figc}).

As we mentioned before, $!$ is a defined combinator.

\begin{definition}[Adjoint, \ensuremath{!~ c}]
 The adjoint of a combinator \ensuremath{c} is defined as follows:

  \begin{itemize}
  \item For primitive isomorphisms \ensuremath{c}, \ensuremath{!~ c} is given by its
    inverse from Figs.~\ref{pi-terms} and~\ref{more-pi}.

  \item \ensuremath{!(c_1 \otimes c_2) =\ !c_1 \otimes~ !c_2}

  \item \ensuremath{!(c_1 \oplus c_2) =\ !c_1 \oplus~ !c_2}

  \item \ensuremath{!(c_1\odot c_2) =\ !c_2 \odot~ !c_1}. (Note that the
    order of combinators has been reversed).

  \end{itemize}
\end{definition}

\noindent We can further define that two combinators are
\emph{observationally equivalent} if on all values of their common
domain, they evaluate to identical values.  More precisely, we will
say that for combinators $c_1, c_2 : b_1 \leftrightarrow b_2$,
$c_1~=~c_2$ whenever:
\[
  \forall  ~v_1:b_1, v_2 : b_2. ~~ c_1 ~v_1 \mapsto v_2 ~\text{if and only if\ } c_2 ~v_1 \mapsto v_2
\]

Each type $b$ has a size $|b|$ defined in the obvious way. We had
previously established that for any natural number $n$, there is a
canonical set of size $n$, which we denoted $[n]$. Furthermore, we can
also define a canonical \emph{type} of that size, which we will denote
$\sharp\, b$, i.e. $\sharp\, b$ is a canonical type of size $|b|$.


\begin{definition}($\sharp$).
  By recursion on $|b|$.  First define $\tau$ that maps numeric sizes
  to their corresponding types. We will revert to using type notation
  for greater clarity of this definition:
\[\begin{array}{rcl}
  \tau~ (0) & = & \bot \\
  \tau~ (1 + n) & = & \top \presumtype \tau~ (n) \\
 \end{array}\]
\noindent so that we can define $\sharp\, b = \tau~ |b|$.
\end{definition}

We are now ready to go further and establish
that there is always an equivalence between a type and the canonical
type of the same size.

\begin{proposition}
  For any type \ensuremath{b} there exists an isomorphism \ensuremath{b \leftrightarrow \sharp\, b}.
  \begin{proof}
    The fact that such an isomorphism exists is evident from the
    definition of size and what it means for two types to be
    isomorphic. While many equivalent constructions are possible
    for any type \ensuremath{b}, one such construction is given by
    \sem{b}:

\[\begin{array}{rclr}
 \sem{0} & = & \idc \\
 \sem{1} & = & \idc \\
 \sem{1{\sumtype}b} & = & \idc \oplus\ \sem{b} \\
 \sem{(b_1{\sumtype}b_2){\sumtype}b_3} & = & \assocrp \odot \sem{b_1 {\sumtype} (b_2 {\sumtype} b_3)} \\
 \sem{b_1 {\sumtype} b_2} & = & (\sem{b_1} \oplus \idc) \odot \sem{ \sharp\, b_1 {\sumtype} b_2 } \\
 \sem{0 {\prodtype} b_2} & = & \absorbr \\
 \sem{1 {\prodtype} b_2} & = & \identlt \odot \sem{b_2} \\
 \sem{(b_1 {\prodtype} b_2) {\prodtype} b_3} & = & \assocrt \odot \sem{b_1 {\prodtype} (b_2 {\prodtype} b_3)} \\
 \sem{(b_1{\sumtype}b_2) {\prodtype} b_3} & = & \dist \odot \sem{b_1 {\prodtype} b_3{\sumtype}b_2 {\prodtype} b_3} \\
 \end{array}\]

  \end{proof}
\end{proposition}

\subsection{Graphical Language}

Combinators of \ensuremath{\Pi } can be written in terms of the
operators described previously or via a graphical language similar in
spirit to those developed for Geometry of Interaction
\cite{DBLP:conf/popl/Mackie95} and string diagrams for category
theory~\cite{BLUTE1996229,selinger-graphical}. Modulo some conventions
and shorthand we describe here, the wiring diagrams are equivalent to
the operator based (syntactic) description of programs.
\ensuremath{\Pi } combinators expressed in this graphical language
look like ``wiring diagrams.'' Values take the form of ``particles''
that flow along the wires. Computation is expressed by the flow of
particles.

\begin{itemize}
\item
The simplest sort of diagram is the \ensuremath{\idc : b \leftrightarrow b} combinator which
is simply represented as a wire labeled by its type \ensuremath{b}. In more
complex diagrams, if the type of a wire is obvious from the context,
it may be omitted.

\inkscape{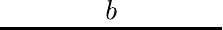}

\noindent
Values flow from left to right in the graphical language of
\ensuremath{\Pi }.  When tracing a computation, one might imagine a value
\ensuremath{v} of type \ensuremath{b} on the wire, as shown below.

\inkscape{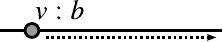}

\item
The product type \ensuremath{b_1 {\prodtype} b_2} may be represented both as one wire
labeled \ensuremath{b_1 \prodtype b_2} or by two parallel wires labeled \ensuremath{b_1} and
\ensuremath{b_2}. Both representations may be used interchangeably.

\begin{multicols}{2}
\inkscape{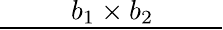}
\inkscape{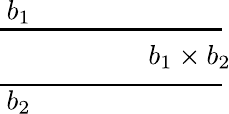}
\end{multicols}

When tracing execution using particles, one should think of one
particle on each wire or alternatively as in folklore in the
literature on monoidal categories as a ``wave.''

\begin{multicols}{2}
\inkscape{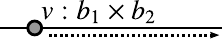}
\inkscape{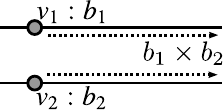}
\end{multicols}

\item
Sum types may similarly be represented using using parallel wires with
a \ensuremath{{\sumtype}} operator between them.

\begin{multicols}{2}
\inkscape{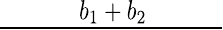}
\inkscape{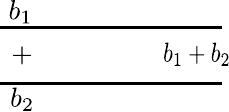}
\end{multicols}

\noindent
When tracing the execution of \ensuremath{b_1{\sumtype}b_2} represented by one
wire, one can think of a value of the form \ensuremath{\mathit{left} ~v_1} or \ensuremath{\mathit{right} ~v_2}
as flowing on the wire, where \ensuremath{v_1:b_1} and \ensuremath{v_2:b_2}.  When tracing
the execution of two additive wires, a value can reside on only one of
the two wires.

\begin{multicols}{2}
\inkscape{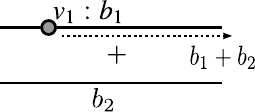}
\inkscape{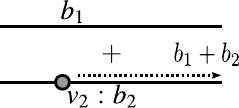}
\end{multicols}

\item
When representing complex types like \ensuremath{(b_1 {\prodtype} b_2){\sumtype}b_3} some visual
grouping of the wires may be done to aid readability. The exact type
however will always be clarified by the context of the diagram.

\inkscape{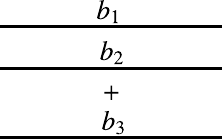}

\item
Associativity is entirely skipped in the graphical language. Hence
three parallel wires may be inferred as \ensuremath{b_1 {\prodtype} (b_2 {\prodtype} b_3)} or
\ensuremath{(b_1 {\prodtype} b_2) {\prodtype} b_3}, based on the context. This is much like handling of
associativity in the graphical representations of categories as well as
that for monoidal categories.

\inkscape{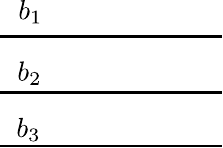}

\item Commutativity is represented by crisscrossing wires.

\begin{multicols}{2}
\inkscape{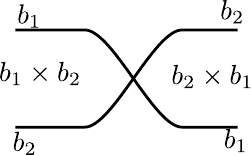}
\inkscape{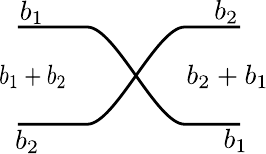}
\end{multicols}

\noindent
When tracing the execution of \ensuremath{b_1{\sumtype}b_2} represented by one wire, one
can think of a value of the form \ensuremath{\mathit{left} ~v_1} or
\ensuremath{\mathit{right} ~v_2} as flowing on the wire, where
\ensuremath{v_1:b_1} and \ensuremath{v_2:b_2}.  By visually
tracking the flow of particles on the wires, one can verify that the
expected types for commutativity are satisfied.

\begin{multicols}{2}
\inkscape{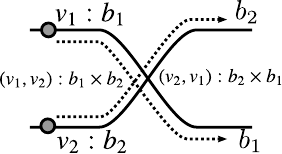}
\inkscape{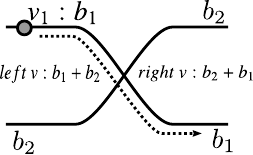}
\end{multicols}

\item
The morphisms that witness that \ensuremath{0} and \ensuremath{1} are the additive and
  multiplicative units are represented as shown below. Note that since there
  is no value of type 0, there can be no particle on a wire of type \ensuremath{0}.
  Also since the monoidal units can be freely introduced and eliminated, sometimes
  they are omitted.  However, as this is in fact dangerous, as explained
  by~\cite{BLUTE1996229}, we will err on the side of including them.

\begin{multicols}{2}
\inkscape{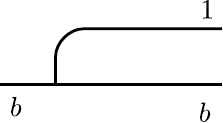}
\inkscape{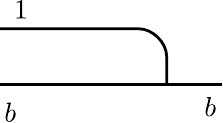}
\end{multicols}

\begin{multicols}{2}
\inkscape{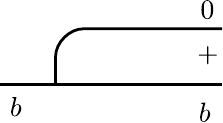}
\inkscape{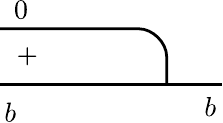}
\end{multicols}

\item
Distributivity and factoring are represented using the dual
boxes shown below:

\begin{multicols}{2}
\inkscape{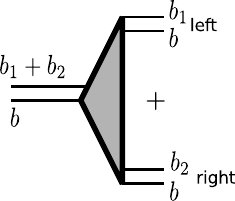}
\inkscape{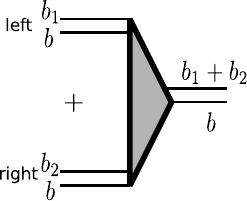}
\end{multicols}

Distributivity and factoring are interesting because they represent
interactions between sum and pair types. Distributivity should
essentially be thought of as a multiplexer that redirects the flow of
\ensuremath{v:b} depending on what value inhabits the type \ensuremath{b_1{\sumtype}b_2}, as shown
below.

\begin{multicols}{2}
\inkscape{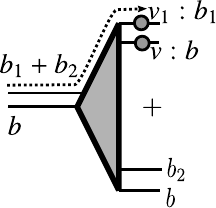}
\inkscape{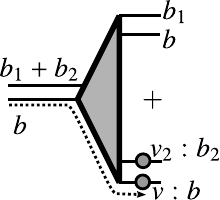}
\end{multicols}

\noindent
Factoring is the corresponding adjoint operation.

\begin{multicols}{2}
\inkscape{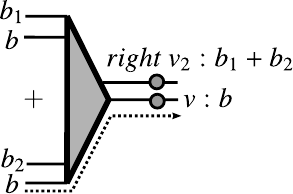}
\inkscape{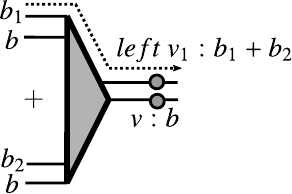}
\end{multicols}

\item Combinators can be composed in series (\ensuremath{c_1 \odot c_2}) or
  parallel. Sequential (series) composition corresponds to connecting
  the output of one combinator to the input of the next.

\inkscape{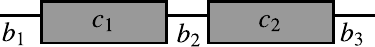}

There are two forms of parallel composition -- combinators
  can be combined additively \ensuremath{c_1 \oplus\ c_2} (shown on the left) or
  multiplicatively \ensuremath{c_1 \otimes c_2} (shown on the right).

\begin{multicols}{2}
\inkscape{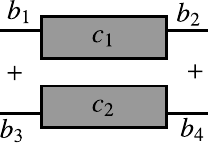}
\inkscape{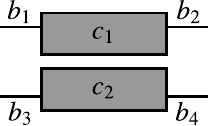}
\end{multicols}

\end{itemize}

\noindent
\textit{Example.} As an example consider the wiring diagram
of the combinator \ensuremath{c} below:
\[\begin{array}{rcl}
c & : & b {\prodtype} (1{\sumtype}1) \leftrightarrow b {\sumtype} b \\
c & = & \swapt \odot \dist \odot (\identlt \oplus \identlt)
\end{array}\]

\begin{center}
\scalebox{1.5}{
\includegraphics{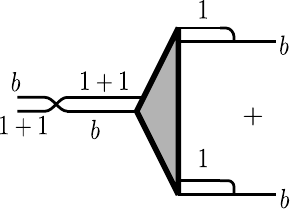}
}
\end{center}

\subsection{Denotational Semantics}

Fig.~\ref{type-isos} introduces our desired denotational semantics,
and Sec.~\ref{sec:opsem} is a direct definition of an operational
semantics. One obvious question arises: do these correspond?

We can certainly associate to each $\Pi$ combinator an
equivalence between the denotation of each type%
\footnote{This is extracted from the Agda formalization
of this work, which has been reported on in a previous paper~\cite{Carette2016}.}:
\begin{code}\hspace*{-4mm}
\>\AgdaFunction{c2equiv} \AgdaSymbol{:} \AgdaSymbol{\{}\AgdaBound{t₁}
\AgdaBound{t₂} \AgdaSymbol{:} \AgdaDatatype{U}\AgdaSymbol{\}} \AgdaSymbol{→}
\AgdaSymbol{(}\AgdaBound{c} \AgdaSymbol{:} \AgdaBound{t₁} \AgdaDatatype{⟷}
\AgdaBound{t₂}\AgdaSymbol{)} \AgdaSymbol{→} \AgdaFunction{⟦} \AgdaBound{t₁}
\AgdaFunction{⟧} \AgdaFunction{≃} \AgdaFunction{⟦} \AgdaBound{t₂}
\AgdaFunction{⟧}%
\end{code}

\noindent And as such an equivalence contains a function as
its first component, we can compare if our operational
semantics and denotational semantics match.  And they do:
\begin{code}\hspace*{-4mm}
\>\AgdaFunction{lemma0} \AgdaSymbol{:} \<[10]%
\>[10]\AgdaSymbol{\{}\AgdaBound{t₁} \AgdaBound{t₂} \AgdaSymbol{:}
\AgdaDatatype{U}\AgdaSymbol{\}} \AgdaSymbol{→} \AgdaSymbol{(}\AgdaBound{c}
\AgdaSymbol{:} \AgdaBound{t₁} \AgdaDatatype{⟷} \AgdaBound{t₂}\AgdaSymbol{)}
\AgdaSymbol{→} \AgdaSymbol{(}\AgdaBound{v} \AgdaSymbol{:} \AgdaFunction{⟦}
\AgdaBound{t₁} \AgdaFunction{⟧}\AgdaSymbol{)} \AgdaSymbol{→}
\AgdaFunction{eval} \AgdaBound{c} \AgdaBound{v} \AgdaDatatype{≡}
\AgdaField{proj₁} \AgdaSymbol{(}\AgdaFunction{c2equiv}
\AgdaBound{c}\AgdaSymbol{)} \AgdaBound{v}\<%
\end{code}

\noindent We can similarly hand-write a backwards evaluator,
prove that it is indeed a proper backwards evaluator, and
finally show that it agrees with the reverse equivalence.

\subsection{Examples}
\label{sec:langRev-examples}
\label{examples}

At first, it is not immediately clear that a programming language
in which information is preserved could model choice. We recall a
quote by Minsky communicating this concern:

\begin{quote}
  Ed Fredkin pursued the idea that information must be finite in
  density. One day, he announced that things must be even more simple
  than that. He said that he was going to assume that information
  itself is conserved. “You’re out of you mind, Ed.” I
  pronounced. “That’s completely ridiculous. Nothing could happen in
  such a world. There couldn’t even be logical gates. No decisions
  could ever be made.” But when Fredkin gets one of his ideas, he’s
  quite immune to objections like that; indeed, they fuel him with
  energy. Soon he went on to assume that information processing must
  also be reversible — and invented what’s now called the Fredkin
  gate~\cite{Hey:1999:FCE:304763}.
\end{quote}

We will however show that one can program all logical gates in
$\Pi$. We will start with a few simple examples and then discuss the
expressiveness of the language and its properties.

\paragraph*{Booleans}
Let us start with encoding booleans. We use the type \ensuremath{1{\sumtype}1} to
represent booleans with \ensuremath{\mathit{left} ~()} representing \ensuremath{\mathit{true}} and
\ensuremath{\mathit{right}~()} representing \ensuremath{\mathit{false}}.
Boolean negation is straightforward to define:

\ensuremath{\mathit{not} : \mathit{bool} \leftrightarrow \mathit{bool}}

\ensuremath{\mathit{not} = \swapp}

\noindent
It is easy to verify that \ensuremath{\mathit{not}} changes \ensuremath{\mathit{true}} to \ensuremath{\mathit{false}} and
vice versa.

\paragraph*{Bit Vectors.}
We can represent $n$-bit words using an n-ary product of
\ensuremath{\mathit{bool}}s. For example, we can represent a 3-bit word, \ensuremath{\mathit{word}_3},
using the type \ensuremath{\mathit{bool} {\prodtype} (\mathit{bool} {\prodtype}  \mathit{bool})}.  We can perform various
operations on these 3-bit words using combinators in \ensuremath{\Pi }. For
instance the bitwise \ensuremath{\mathit{not}} operation is the parallel composition of
three \ensuremath{\mathit{not}} operations:

\ensuremath{\mathit{not}_{\mathit{word}_3} :: \mathit{word}_3 \leftrightarrow \mathit{word}_3}

\ensuremath{\mathit{not}_{\mathit{word}_3} = \mathit{not}  {\prodtype}  (\mathit{not}  {\prodtype}  \mathit{not})}

\noindent We can express a 3-bit word reversal operation as follows:

\ensuremath{\mathit{reverse} : \mathit{word}_3 \leftrightarrow \mathit{word}_3}

\ensuremath{\mathit{reverse} = \swapt \odot (\swapt  \otimes  \idc)~ \odot \assocrt}

\noindent We can check that \ensuremath{\mathit{reverse}} does the right thing by
applying it to a value \ensuremath{(v_1, (v_2, v_3))} and writing out the full
derivation tree of the reduction.  The combinator \ensuremath{\mathit{reverse}}, like
many others we will see in this paper, is formed by sequentially
composing several simpler combinators. Instead of presenting the
operation of \ensuremath{\mathit{reverse}} as a derivation tree, it is easier (purely
for presentation reasons) to flatten the tree into a sequence of
reductions as caused by each component. Such a sequence of reductions
is given below:
\[\begin{array}{rlr}
 & (v_1, (v_2, v_3)) \\
 \swapt & ((v_2, v_3), v_1) \\
 \swapt \otimes  \idc & ((v_3, v_2), v_1) \\
 \assocrt & (v_3, (v_2, v_1)) \\
 \end{array}\]

\noindent On the first line is the initial value. On each subsequent
line is a fragment of the \ensuremath{\mathit{reverse}} combinator and the value that
results from applying this combinator to the value on the previous
line. For example, \ensuremath{\swapt} transforms \ensuremath{(v_1, (v_2, v_3))} to
\ensuremath{((v_2,v_3),v_1)}.  On the last line we see the expected result with
the bits in reverse order.

We can also draw out the graphical representation of the 3-bit reverse
combinator. In the graphical representation, it is clear that the
combinator achieves the required shuffling.

\inkscape{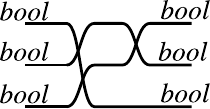}

\paragraph*{Conditionals.}
Even though \ensuremath{\Pi } lacks conditional expressions, they are
expressible using the distributivity and factoring laws. The
diagrammatic representation of \ensuremath{\dist} shows that it redirects the flow
of a value \ensuremath{v:b} based on the value of another one of type
\ensuremath{b_1{\sumtype}b_2}. If we choose \ensuremath{1{\sumtype}1} to be
\ensuremath{\mathit{bool}} and apply either \ensuremath{c_1:b_1\leftrightarrow
b_2} or \ensuremath{c_2:b_1\leftrightarrow b_2} to the value \ensuremath{v},
then we essentially have an `if' expression.

\ensuremath{\mathit{if}_{c_1,c_2} : \mathit{bool}  {\prodtype}  b_1 \leftrightarrow \mathit{bool}  {\prodtype}  b_2}

\ensuremath{\mathit{if}_{c_1,c_2} = \dist \odot ((\idc  \otimes\  c_1) {\sumtype} (\idc \otimes\  c_2)) \odot \factor}

\inkscape{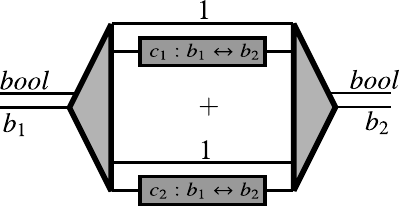}

The diagram above shows the input value of type \ensuremath{(1{\sumtype}1) {\prodtype}  b_1}
processed by the distribute operator \ensuremath{\dist}, which converts it into
a value of type \ensuremath{(1 {\prodtype}  b_1){\sumtype}(1 {\prodtype}  b_1)}. In the
\ensuremath{\mathit{left}} branch, which corresponds to the
case when the boolean is \ensuremath{\mathit{true}} (i.e. the value was
\ensuremath{\mathit{left} ~()}), the combinator~\ensuremath{c_1} is applied to
the value of type~\ensuremath{b_1}. The right
branch which corresponds to the boolean being \ensuremath{\mathit{false}} passes
the value of type \ensuremath{b_1} through the combinator \ensuremath{c_2}.
The inverse of \ensuremath{\dist}, namely \ensuremath{\factor} is applied
to get the final result of type \ensuremath{(1{\sumtype}1) {\prodtype} b_2}.

\paragraph*{Logic Gates}
There are several universal primitives for conventional (irreversible)
hardware circuits, such as \ensuremath{\mathit{nand}} and \ensuremath{\mathit{fanout}}. In the case
of reversible hardware circuits, the canonical universal primitive is
the Toffoli gate~\cite{Toffoli:1980}. The Toffoli gate takes three
boolean inputs: if the first two inputs are \ensuremath{\mathit{true}} then the third
bit is negated. In a traditional language, the Toffoli gate would be
most conveniently expressed as a conditional expression like:

\noindent
\ensuremath{ \mathit{toffoli}(v_1,v_2,v_3) = \mathit{if} ~(v_1 ~\mathit{and} ~v_2) ~\mathit{then} ~(v_1, v_2, \mathit{not}(v_3)) ~\mathit{else} ~(v_1, v_2, v_3)}

We will derive Toffoli gate in \ensuremath{\Pi } by first deriving a simpler
logic gate called \ensuremath{\mathit{cnot}}.  Consider a one-armed version, \ensuremath{\mathit{if}_c},
of the conditional derived above. If the \ensuremath{\mathit{bool}} is
\ensuremath{\mathit{true}}, the value of type \ensuremath{b} is modified by the operator \ensuremath{c}.

\begin{center}
\scalebox{1.5}{
\includegraphics{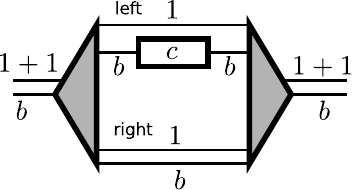}
}
\end{center}

By choosing \ensuremath{b} to be \ensuremath{\mathit{bool}} and \ensuremath{c} to be \ensuremath{\mathit{not}}, we have the
combinator \ensuremath{\mathit{if}_{\mathit{not}} : \mathit{bool} {\prodtype}  \mathit{bool}\leftrightarrow \mathit{bool} {\prodtype}  \mathit{bool}} which negates its
second argument if the first argument is \ensuremath{\mathit{true}}. This gate
\ensuremath{\mathit{if}_{\mathit{not}}} is often referred to as the \ensuremath{\mathit{cnot}} gate\cite{Toffoli:1980}.

If we iterate this construction once more, the resulting combinator
\ensuremath{\mathit{if}_{\mathit{cnot}}} has type \ensuremath{\mathit{bool} {\prodtype}  (\mathit{bool} {\prodtype}  \mathit{bool})\leftrightarrow \mathit{bool} {\prodtype}  (\mathit{bool} {\prodtype}  \mathit{bool})}. The
resulting gate checks the first argument and if it is \ensuremath{\mathit{true}},
proceeds to check the second argument. If that is also \ensuremath{\mathit{true}} then
it will negate the third argument. Thus \ensuremath{\mathit{if}_{\mathit{cnot}}} is the required
Toffoli gate.

\begin{center}
\scalebox{1.6}{
\includegraphics{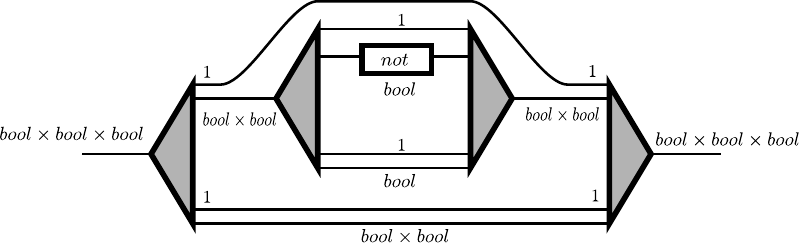}
}
\end{center}

\section{Data III: Reversible Programs between Reversible Programs}
\label{sec:pi2}

In the previous sections, we examined equivalences between
conventional data structures, i.e., sets of values and structured trees
of values. We now consider a richer but
foundational notion of data: programs themselves. Indeed, universal
computation models crucially rely on the fact that \emph{programs
are (or can be encoded as) data}, e.g., a Turing machine can be
encoded as a string that another Turing machine (or even the same
machine) can manipulate. Similarly, first-class functions are
the \emph{only} values in the $\lambda$-calculus.
In our setting, the programs developed in the
previous section are reversible deformations between structured finite
types. We now ask whether these programs can themselves
be subject to (higher-level) reversible deformations?

Before developing the theory, let's consider a small example
consisting of two deformations between the types $A + B$ and $C+D$:

\begin{center}
\begin{tikzpicture}[scale=0.7,every node/.style={scale=0.8}]
  \draw[>=latex,<->,double,red,thick] (2.25,-1.2) -- (2.25,-2.9) ;
  \draw[fill] (-2,-1.5) circle [radius=0.025];
  \node[below] at (-2.1,-1.5) {$A$};
  \node[below] at (-2.1,-1.9) {$+$};
  \draw[fill] (-2,-2.5) circle [radius=0.025];
  \node[below] at (-2.1,-2.5) {$B$};

  \draw[fill] (6.5,-1.5) circle [radius=0.025];
  \node[below] at (6.7,-1.5) {$C$};
  \node[below] at (6.7,-1.9) {$+$};
  \draw[fill] (6.5,-2.5) circle [radius=0.025];
  \node[below] at (6.7,-2.5) {$D$};

  \draw[<-] (-2,-1.5) to[bend left] (1,0.5) ;
  \draw[<-] (-2,-2.5) to[bend left] (1,-0.5) ;
  \draw[->] (3.5,0.5) to[bend left] (6.5,-1.45) ;
  \draw[->] (3.5,-0.5) to[bend left] (6.5,-2.45) ;

  \draw[<-] (-2,-1.5) to[bend right] (1,-3.5) ;
  \draw[<-] (-2,-2.5) to[bend right] (1,-4.5) ;
  \draw[->] (3.5,-3.5) to[bend right] (6.5,-1.55) ;
  \draw[->] (3.5,-4.5) to[bend right] (6.5,-2.55) ;

  \draw     (2,0.5)  -- (2.5,0.5)  ;
  \draw     (2,-0.5) -- (2.5,-0.5) ;

  \draw     (2.5,0.5)  -- (3.5,-0.5)  ;
  \draw     (2.5,-0.5) -- (3.5,0.5) ;

  \draw     (1,-3.5)  -- (2,-4.5)    ;
  \draw     (1,-4.5) -- (2,-3.5)   ;

  \draw     (2,-3.5)  -- (2.5,-3.5)    ;
  \draw     (2,-4.5) -- (2.5,-4.5)   ;

  \path (1.5,0.5) node (tc1) [func] {$c_1$};
  \path (1.5,-0.5) node (tc2) [func] {$c_2$};
  \path (3,-4.5) node (bc1) [func] {$c_1$};
  \path (3,-3.5) node (bc2) [func] {$c_2$};
\end{tikzpicture}
\end{center}
The top path is the $\Pi$ program
$(c_1~\oplus~c_2)~\odot~\swapp$ which deforms the
type $A$ by $c_1$, deforms the type $B$ by $c_2$, and deforms the
resulting space by a twist that exchanges the two injections into the
sum type. The bottom path performs the twist first and then deforms
the type $A$ by $c_1$ and the type $B$ by $c_2$ as before. One
could imagine the paths are physical \emph{elastic} wires in $3$ space, where
the deformations $c_1$ and $c_2$ as arbitrary deformations on these wires, and
the twists do not touch but are in fact well-separated. Then, holding the
points $A$, $B$, $C$, and $D$ fixed, it is possible to imagine
sliding $c_1$ and $c_2$ from the top wire rightward past the
twist, and then using the elasticity of the wires, pull the
twist back to line up with that of the bottom --- thus making
both parts of the diagram identical.  Each of these moves
can be undone (reversed), and doing so would take the bottom
part of the diagram into the top part.  In other
words, there exists a deformation of the program
$(c_1~\oplus~c_2)~\odot~\swapp$ to the program
$\swapp \odot (c_2~\oplus~c_1)$. We can also show that this
means that, as permutations, $(c_1~\oplus~c_2)~\odot~\swapp$ and
$\swapp \odot (c_2~\oplus~c_1)$ are equal. And, of course, not
all programs between the same types can be deformed into one
another. The simplest example of inequivalent deformations
are the two automorphisms of $1+1$, namely $\idc$ and $\swapp$.

While we will not make the details of the stretchable wires and
slidable boxes formal, it is useful for intuition.  One caveat
though: some of the sliding and stretching needs to be done in
spaces of higher dimension than 3 to have ``enough room'' to
move things along without collision or over-stretching wires.
That, unfortunately, means that some equivalences are harder to
grasp. Luckily, most equivalences only need 3 dimensions.

Our reversible language of type isomorphisms and equivalences between
them has a strong connection to \emph{univalent universes} in
HoTT~\cite{Carette2018}. Based on this connection, we refer to the
types as being at level-0, to the equivalences between types (i.e., the
combinators of Sec.~\ref{sec:pi1}) as being at level-1, and to the
equivalences between equivalences of types (i.e., the combinators
discussed in this section) as being at level-2.

\subsection{A Model of Equivalences between Type Equivalences}

Previously we saw how we could take the proof terms of commutative semiring
equivalences as our starting point for $\Pi$. What we need
now is to understand how \emph{proofs} of algebraic identities should be
considered equivalent. Classical algebra does not help, as proofs
are not considered first-class citizens. However,
another route is available to us: since the work of
Hofmann and Streicher~\cite{hofmann96thegroupoid}, we know that
one can model types as \emph{groupoids}.  The additional
structure comes from explicitly modeling the ``identity
types'': instead of regarding all terms which witness
the equality of (say) $a$ and $b$ of type $A$ as being
indistinguishable, we posit that there may in fact be many.
This consequences of this one decision are enough to show that
types can be modeled by groupoids.

Thus, rather than looking at (untyped) commutative semirings, we
should look at a \emph{typed} version. This process frequently goes by
the moniker of ``categorification.''  We want a categorical algebra,
where the basic objects are groupoids (to model our types), and where
there is a natural notion of $+$ and $*$.  At first, we hit what seems
like a serious stumbling block: the category of all groupoids, \Gpd,
have neither co-products nor products. However, we don't want to work
internally in \Gpd -- we want operations \emph{on} groupoids. In other
words, we want something akin to symmetric monoidal categories, but
with two interacting monoidal structures.  Luckily, this already
exists: the categorical analog to (commutative) semirings are
(symmetric) Rig Categories~\cite{laplaza72,kelly74}.  This
straightforwardly generalizes to symmetric Rig Groupoids.

How does this help? Coherence conditions! Symmetric monoidal categories,
to start somewhere simple, do not just introduce natural transformations
like the associator $\alpha$ and the left and right unitors ($\lambda$
and~$\rho$ respectively), but also coherence conditions that these must satisfy.
Looking, for example, at just the additive fragment of $\Pi$ (i.e. with just $0$,
$1$ and $+$ for the types, $\odot$ and $\oplus$ as combinators, and
only the terms so expressible), the sub-language would correspond,
denotationally, to exactly (non-empty) symmetric monoidal groupoids. And what
these possess are exactly some \emph{equations between equations}
as commutative diagrams.  Transporting these coherence conditions, for
example those that express that various transformations are \emph{natural}
to $\Pi$, gives a list of equations between $\Pi$ programs.
Furthermore, all the natural transformations
that arise are in fact natural \emph{isomorphisms} -- and thus
reversible.

We can then proceed to prove that every one of the coherence conditions
involved in defining a symmetric Rig Groupoid holds for the groupoid
interpretation of types~\cite{Carette2016}.  This is somewhat tedious
given the sheer number of these, but when properly formulated,
relatively straightforward, but see below for comments on some
tricky cases.

But why are these particular coherence laws? Are they all necessary?
Conversely are they, in some appropriate sense, sufficient? This is
the so-called \emph{coherence problem}. Mac Lane, in his farewell address
as President of the American Mathematical Society~\cite{MacLane1976} gives
a good introduction and overview of such problems.  A more modern
interpretation (which can nevertheless be read into Mac Lane's own
exposition) would read as follows: given a set of equalities on abstract
words, regarded as a rewrite system, and two means of rewriting a word
in that language to another, is there some suitable notion of canonical
form that expresses the essential uniqueness of the non-trivial
rewrites?  Note how this word-and-rewrite problem is essentially
independent of the eventual interpretation. But one must take some care,
as there are obvious degenerate cases (involving ``trivial'' equations
involving $0$ or $1$) which lead to non-uniqueness. The landmark
results, first by Kelly-Mac Lane~\cite{KELLY197197} for closed
symmetric monoidal categories, then (independently) Laplaza and
Kelly~\cite{laplaza72,kelly74} for symmetric Rig Categories, is
that indeed there are sound and complete coherence conditions that
insure that all the ``obvious'' equalities between different abstract
words in these systems give rise to commutative diagrams. The
``obvious'' equalities come from \emph{syzygies} or
\emph{critical pairs} of the system of equations.
The problem is far from trivial --- Fiore et al.~\cite{Fiore-2008}
document some publications where the coherence set is in
fact incorrect. They furthermore give a quite general algorithm
to derive such coherence conditions.

\subsection{A Language of Equivalences between Type Equivalences}
\label{langeqeq}

As motivated in the previous section, the equivalences between type
equivalences are perfectly modeled by the coherence conditions of weak
Rig Groupoids. Syntactically, we take the easiest way there: simply
make every coherence isomorphism into a programming construct. These
constructs are collected in several figures (Fig.~\ref{figj} to
Fig.~\ref{figa}) and are discussed next.

Conveniently, the various coherence conditions can be naturally
grouped into ``related'' laws.  Each group basically captures the
interactions between compositions of level-1 $\Pi$ combinators.

\begin{figure}[t]
Let $c_1 : t_1 \leftrightarrow t_2$, $c_2 : t_3 \leftrightarrow t_4$, $c_3 : t_1 \leftrightarrow t_2$, and $c_4 : t_3 \leftrightarrow t_4$. \\
Let $a_1 : t_5 \leftrightarrow t_1$,  $a_2 : t_6 \leftrightarrow t_2$, $a_3 : t_1 \leftrightarrow t_3$, and $a_4 : t_2 \leftrightarrow t_4$.
\[\def\arraystretch{1.3}
\begin{array}{c}
\Rule{}
  {c_1 \Leftrightarrow c_3 \quad c_2 \Leftrightarrow c_4}
  {c_1 \oplus c_2 \Leftrightarrow c_3 \oplus c_4}
  {}
\qquad
\Rule{}
  {c_1 \Leftrightarrow c_3 \quad c_2 \Leftrightarrow c_4}
  {c_1 \otimes c_2 \Leftrightarrow c_3 \otimes c_4}
  {}
\\
  {(a_1 \odot a_3) \oplus (a_2 \odot a_4) \Leftrightarrow (a_1 \oplus a_2) \odot (a_3 \oplus a_4)}
\\
  {(a_1 \odot a_3) \otimes (a_2 \odot a_4) \Leftrightarrow (a_1 \otimes a_2) \odot (a_3 \otimes a_4)}
\end{array}\]
\caption{\label{fige}Signatures of level-2 $\Pi$-combinators: functors}
\end{figure}

Starting with the simplest constructions, the first two constructs in
Fig.~\ref{fige} are the level-2 analogs of~$+$ and~$*$, which
respectively model level-1 choice composition and parallel composition
(of equivalences).  These allow us to ``build up'' larger equivalences
from smaller ones.  The next two express that both of these
composition operators distribute over sequential composition $\odot$
(and vice versa).

\begin{figure}[t]
Let $c_1 : t_1 \leftrightarrow t_2$,  $c_2 : t_2 \leftrightarrow t_3$, and $c_3 : t_3 \leftrightarrow t_4$:
\[\def\arraystretch{1.3}
\begin{array}{c}
  {c_1 \odot (c_2 \odot c_3) \Leftrightarrow (c_1 \odot c_2) \odot c_3}
\\
  {(c_1 \oplus (c_2 \oplus c_3)) \odot \assoclp \Leftrightarrow \assoclp \odot ((c_1 \oplus c_2) \oplus c_3)}
\\
  {(c_1 \otimes (c_2 \otimes c_3)) \odot \assoclt \Leftrightarrow \assoclt \odot ((c_1 \otimes c_2) \otimes c_3)}
\\
  {((c_1 \oplus c_2) \oplus c_3) \odot \assocrp \Leftrightarrow \assocrp \odot (c_1 \oplus (c_2 \oplus c_3))}
\\
  {((c_1 \otimes c_2) \otimes c_3) \odot \assocrt \Leftrightarrow \assocrt \odot (c_1 \otimes (c_2 \otimes c_3))}
\\
  {\assocrp \odot \assocrp \Leftrightarrow ((\assocrp \oplus \idc) \odot \assocrp) \odot (\idc \oplus \assocrp)}
\\
  {\assocrt \odot \assocrt \Leftrightarrow ((\assocrt \otimes \idc) \odot \assocrt) \odot (\idc \otimes \assocrt)}
\end{array}\]
\caption{\label{figj}Signatures of level-2 $\Pi$-combinators: associativity}
\end{figure}

The constructs in Fig.~\ref{figj} capture the informal idea that all
the different ways of associating programs are equivalent. The first
says that sequential composition itself ($\odot$) is associative.
The next $4$ capture how
the $\oplus$ and $\otimes$ combinators ``commute'' with re-association.
In other words, it expresses that the type-level associativity of $+$ is
properly reflected by the properties of $\oplus$.
The last two equivalences show how composition of associativity combinators
interact together.

The bottom line in Fig.~\ref{figj} is actually a linear
restatement of the famous ``pentagon diagram'' stating a
particular coherence condition for monoidal categories~\cite{KELLY197197}.
To make the relation between $\Pi$ as a language and the
language of category theory, the figure below displays
the same morphism but in categorical terms.

\begin{center}
\begin{tikzcd}[column sep=normal]
   & (A \times (B \times C)) \times D \arrow [dr, "\assocrt"] & \\
((A \times B) \times C) \times D \arrow [ur, "\assocrt \otimes \mathit{id}\leftrightarrow"]
   \arrow [d, "\assocrt"] &
       & A \times ((B \times C) \times D) \arrow [d, "\mathit{id}\leftrightarrow \otimes \assocrt" ]\\
(A \times B) \times (C \times D) \arrow [rr, "\assocrt"] & & A \times (B \times (C \times D))
\end{tikzcd}
\end{center}

\begin{figure}[t]
Let $c_1 : t_1 \leftrightarrow t_2$, $c_2 : t_3 \leftrightarrow t_4$, and $c_3 : t_5 \leftrightarrow t_6$:
\[\def\arraystretch{1.3}
\begin{array}{c}
  {((c_1 \oplus c_2) \otimes c_3) \odot \dist \Leftrightarrow \dist \odot ((c_1 \otimes c_3) \oplus (c_2 \otimes c_3))}
\\
  {(c_1 \otimes (c_2 \oplus c_3)) \odot \distl \Leftrightarrow \distl \odot ((c_1 \otimes c_2) \oplus (c_1 \otimes c_3))}
\\
  {((c_1 \otimes c_3) \oplus (c_2 \otimes c_3)) \odot \factor \Leftrightarrow \factor \odot ((c_1 \oplus c_2) \otimes c_3)}
\\
  {((c_1 \otimes c_2) \oplus (c_1 \otimes c_3)) \odot \factorl \Leftrightarrow \factorl \odot (c_1 \otimes (c_2 \oplus c_3))}
\end{array}\]
\caption{\label{figi}Signatures of level-2 $\Pi$-combinators: distributivity and factoring}
\end{figure}

The constructs in Fig.~\ref{figi} are the basic coherence for
$\dist$, $\distl$, $\factor$ and $\factorl$: the type-level distribution
and factoring has to commute with the level-1 $\oplus$ and $\otimes$.

\begin{figure}[t]
Let $c_0, c_1, c_2, c_3 : t_1 \leftrightarrow t_2$ and $c_4, c_5 : t_3 \leftrightarrow t_4$:
\[\def\arraystretch{1.3}
\begin{array}{c}
  {\idc \odot \, c_0 \Leftrightarrow c_0}
\quad
  {c_0 \, \odot \idc \, \Leftrightarrow c_0}
\quad
  {c_0\,\, \odot\,!\, c_0 \Leftrightarrow \idc}
\quad
  {!\, c_0 \odot c_0 \Leftrightarrow \idc}
\\
  {\idc \oplus \, \idc \, \Leftrightarrow \idc}
\qquad
  {\idc \otimes \, \idc \, \Leftrightarrow \idc}
\\
  {c_0 \Leftrightarrow c_0}
\quad
\Rule{}
  {c_1 \Leftrightarrow c_2 \quad c_2 \Leftrightarrow c_3}
  {c_1 \Leftrightarrow c_3}
  {}
\quad
\Rule{}
  {c_1 \Leftrightarrow c_4 \quad c_2 \Leftrightarrow c_5}
  {c_1 \odot c_2 \Leftrightarrow c_4 \odot c_5}
  {}
\end{array}\]
\caption{\label{figh}Signatures of level-2 $\Pi$-combinators: identity and composition}
\end{figure}

The constructs in Fig.~\ref{figh} express various properties of composition.
The first two says that $\idc$ is a left and right identity for sequential composition.
The next two say that all programs are reversible, both on the left and the right:
running $c$ and then its reverse ($!\, c$) is equivalent to the identity, and the
same for doing $!\, c$ first then $c$. The last line say that there is an
identity level-2 combinator, a sequential composition, and that level-2
equivalence respects level-1 sequential composition $\odot$.

\begin{figure}[t]
Let $c_0 : 0 \leftrightarrow 0$, $c_1 : 1 \leftrightarrow 1$, and $c_3 : t_1 \leftrightarrow t_2$:
\[\def\arraystretch{1.3}
\begin{array}{c}
  {\identlp \odot c_3 \Leftrightarrow (c_0 \oplus c_3) \odot \identlp}
\qquad
  {\identrp \odot (c_0 \oplus c_3) \Leftrightarrow c_3 \odot \identrp}
\\
  {\identlsp \odot c_3 \Leftrightarrow (c_3 \oplus c_0) \odot \identlsp}
\qquad
  {\identrsp \odot (c_3 \oplus c_0) \Leftrightarrow c_3 \odot \identrsp}
\\
  {\identlt \odot c_3 \Leftrightarrow (c_1 \otimes c_3) \odot \identlt}
\qquad
  {\identrt \odot (c_1 \otimes c_3) \Leftrightarrow c_3 \odot \identrp}
\\
  {\identlst \odot c_3 \Leftrightarrow (c_3 \otimes c_1) \odot \identlst}
\qquad
  {\identrst \odot (c_3 \otimes c_1) \Leftrightarrow c_3 \odot \identrst}
\\
  {\identlt \Leftrightarrow \distl \odot (\identlt \oplus \identlt)}
\\
\identlp \Leftrightarrow \swapp \odot \identlsp
\qquad
\identlt \Leftrightarrow \swapt \odot \identlst
\end{array}\]
\caption{\label{figg}Signatures of level-2 $\Pi$-combinators: unit}
\end{figure}

The constructs in Fig.~\ref{figg} may at first blush look similarly straightforward,
but deserve some pause. One obvious question: What is the point of
$c_0 : 0 \leftrightarrow 0$, isn't that just the identity combinator $\idc$
for $A = 0$ (as defined in Fig.~\ref{type-isos})? Operationally, $c_0$
is indeed indistinguishable from $\idc$. However, there are multiple syntactic
ways of writing down combinators of type $0 \leftrightarrow 0$, and the
first combinator in Fig.~\ref{figg} applies to all of them uniformly.
This is another subtle aspect of coherence: all reasoning must be valid for
all possible models, not just the one we have in mind. So even though
operational reasoning may suggest that some relations \emph{may} be
true between combinators, it can also mislead. The same reasoning
applies to $c_1 : 1 \leftrightarrow 1$.  The first $8$ combinators can
then be read as basic coherence for unit introduction and elimination,
in both additive and multiplicative cases.

The last two capture
another simple idea, related to swapping: eliminating a unit
on the left is the same as first swapping then eliminating on the
right (both additively and multiplicatively). As a side note,
these are not related to \emph{commutativity}, but rather
come from one of the simplest coherence condition for
braided monoidal categories. In other words, it reflects the
idempotence of $\swapp$ and $\swapt$ rather than the
commutativity of $\oplus$ and $\otimes$.

\begin{figure}[t]
Let $c_1 : t_1 \leftrightarrow t_2$ and $c_2 : t_3 \leftrightarrow t_4$:
\[\def\arraystretch{1.3}
\begin{array}{c}
  {\swapp \odot (c_1 \oplus c_2) \Leftrightarrow (c_2 \oplus c_1) \odot \swapp}
\quad
  {\swapt \odot (c_1 \otimes c_2) \Leftrightarrow (c_2 \otimes c_1) \odot \swapt}
\\
  {(\assocrp \odot \swapp) \odot \assocrp \Leftrightarrow ((\swapp \oplus \idc) \odot \assocrp) \odot (\idc \oplus \swapp)}
\\
  {(\assoclp \odot \swapp) \odot \assoclp \Leftrightarrow ((\idc \oplus \swapp) \odot \assoclp) \odot (\swapp \oplus \idc)}
\\
  {(\assocrt \odot \swapt) \odot \assocrt \Leftrightarrow ((\swapt \otimes \idc) \odot \assocrt) \odot (\idc \otimes \swapt)}
\\
  {(\assoclt \odot \swapt) \odot \assoclt \Leftrightarrow ((\idc \otimes \swapt) \odot \assoclt) \odot (\swapt \otimes \idc)}
\end{array}\]
\caption{\label{figf}Signatures of level-2 $\Pi$-combinators: commutativity and associativity}
\end{figure}

The first two equivalences in Fig.~\ref{figf} reflect the basic
coherence between level-0 swapping and the level-1 combinator
actions. The next four arise because of interactions between (additive
and multiplicative) level-1 associativity and swapping.  In other
words, they arise as critical pairs.  For example, the first expresses
that the two ways of going from $\left(A \oplus B\right) \oplus C$ to
$B \oplus \left(C \oplus A\right)$ are equivalent, with the second
saying that the reverse (i.e.  the results of applying $!$\,) also
gives equivalent programs.  The last two say the same but for the
multiplicative structure.

\begin{figure}[t]
\[\def\arraystretch{1.3}
\begin{array}{c}
  {\identlsp \oplus \idc ~\Leftrightarrow~ \assocrp \odot (\idc \oplus \, \identlp)}
\\
  {\identlst \otimes \idc ~\Leftrightarrow~ \assocrt \odot (\idc \otimes \, \identlt)}
\end{array}\]
\caption{\label{figd}Signatures of level-2 $\Pi$-combinators: unit and associativity}
\end{figure}

The constructs in Fig.~\ref{figd} express how unit elimination ``in the middle''
can be expressed either as operating on the right or, (after re-association) on the left.

\begin{figure}[t]
Let $c : t_1 \leftrightarrow t_2$:
\[\def\arraystretch{1.3}
\begin{array}{c}
  {(c \otimes \idc) \odot \absorbl \Leftrightarrow \absorbl \odot \idc}
\quad
  {(\idc \, \otimes c) \odot \absorbr \Leftrightarrow \absorbr \odot \idc}
\\
  {\idc \odot \, \factorzl \Leftrightarrow \factorzl \odot (\idc \otimes c)}
\quad
  {\idc \odot \, \factorzr \Leftrightarrow \factorzr \odot (c \otimes \idc)}
\\
  {\absorbr \Leftrightarrow \absorbl}
\\
  {\absorbr \Leftrightarrow (\distl \odot (\absorbr \oplus \absorbr)) \odot \identlp}
\\
  {\identlst \Leftrightarrow \absorbr}
\qquad
  {\absorbl \Leftrightarrow \swapt \odot \absorbr}
\\
  {\absorbr \Leftrightarrow (\assoclt \odot (\absorbr \otimes \idc)) \odot \absorbr}
\\
  {(\idc \otimes \absorbr) \odot \absorbl \Leftrightarrow (\assoclt \odot (\absorbl \otimes \idc)) \odot \absorbr}
\\
  {\idc \otimes \, \identlp \Leftrightarrow (\distl \odot (\absorbl \oplus \idc)) \odot \identlp}
\end{array}\]
\caption{\label{figc}Signatures of level-2 $\Pi$-combinators: zero}
\end{figure}

The constructs in Fig.~\ref{figc} are significantly more subtle, as they
deal with combinators involving $0$, aka an impossibility.  For example,
\[  {(c \otimes \idc_{0}) \odot \absorbl \Leftrightarrow \absorbl \odot \idc_{0}}
\]
(where we have explicitly annotated the types of $\idc$ for increased clarity)
tells us that of the two ways of transforming from $t_1  *\  0$ to $0$,
namely first doing some arbitrary transformation $c$ from $t_1$ to $t_2$ and
(in parallel) leaving $0$ alone then eliminating $0$, or first eliminating $0$
then doing the identity (at $0$), are equivalent. This is the ``naturality'' of
$\absorbl$. One item to note is the fact that this combinator is not
irreducible, as the $\idc$ on the right can be eliminated. But that is actually
a property visible at an even higher level (which we will not touch in this
paper).  The next $3$ are similarly expressing the naturality of $\absorbr$,
$\factorzl$ and $\factorzr$.

The next combinator, $\absorbr \Leftrightarrow \absorbl$, is
particularly fascinating: while it says something simple --- that the
two obvious ways of transforming $0 * 0$ into $0$, namely absorbing
either the left or right $0$ --- it implies something subtle.  A
straightforward proof of $\absorbl$ which proceeds by saying that
$0 * t$ cannot be inhabited because the first member of the pair
cannot, is not in fact equivalent to $\absorbr$ on $0 * 0$.  However,
if we instead define $\absorbl$ to ``transport'' the putative
impossible first member of the pair to its (equally impossible)
output, then these do form equivalent pairs.  The next few in
Fig.~\ref{figc} also express how $\absorbr$ and $\absorbl$ interact
with other combinators. As seen previously, all of these arise as
critical pairs. What is much more subtle here is that the types
involved often are asymmetric: they do not have the same occurrences
on the left and right. Such cases are particularly troublesome for
finding normal forms. Laplaza~\cite{laplaza72} certainly comments on this,
but in mostly terse and technical terms. Blute et al.~\cite{BLUTE1996229}
offer much more intuitive explanations.

\begin{figure}[t]
\[\def\arraystretch{1.3}
\begin{array}{c}
  {((\assoclp \otimes \idc) \odot \dist) \odot (\dist \oplus \idc) \Leftrightarrow (\dist \odot (\idc \oplus \dist)) \odot \assoclp}
\\
  {\assoclt \odot \distl \Leftrightarrow ((\idc \otimes \distl) \odot \distl) \odot (\assoclt \oplus \assoclt)}
\end{array}\]
\vspace{ -0.5em}
\[\def\arraystretch{1.3}
\begin{array}{rcl}
  (\distl \odot (\dist \oplus \dist)) \odot \assoclp &\Leftrightarrow&
   \dist \odot (\distl \oplus \distl) \odot \assoclp ~\odot \\
&& (\assocrp \oplus \idc) ~\odot \\
&& ((\idc \oplus \swapp) \oplus \idc) ~\odot \\
&&      (\assoclp \oplus \idc)
\end{array}\]
\caption{\label{figb}Signatures of level-2 $\Pi$-combinators: associativity and distributivity}
\end{figure}

\begin{figure}[t]
\[\def\arraystretch{1.3}
\begin{array}{rcl}
  (\idc \otimes \swapp) \odot \distl &\Leftrightarrow& \distl \odot \swapp
\\
  \dist \odot (\swapt \oplus \swapt) &\Leftrightarrow & \swapt \odot \distl
\end{array}\]
\caption{\label{figa}Signatures of level-2 $\Pi$-combinators: commutativity and distributivity}
\end{figure}

The constructs in Fig.~\ref{figb} and Fig.~\ref{figa} relating associativity and
distributivity, and commutativity and distributivity, have more in common with
previous sets of combinators.  They do arise from non-trivial critical pairs
of different ways of going between equivalent types. The last one of
Fig.~\ref{figb} is particularly daunting, involving a sequence of $3$ combinators
on the left and $6$ on the right.

\subsection{Operational Semantics}

There are two different interpretations for an operational semantics
for the language of equivalences:
\begin{enumerate}
\item Mimicking closely the one in Sec.~\ref{sec:opsem}, and thus
  finding explicit homotopies between the functions induced by the
  operational semantics of the level-1 combinators.
\item Treating things more syntactically, and interpreting
the combinators as program transformations.
\end{enumerate}
A previous paper~\cite{Carette2016} explores the first interpretation
in depth. There one can find a definition of ``equivalences of
equivalences'', which as the base of that interpretation.

\newcommand{\evalone}{\ensuremath{\mathit{eval}_1}}

Here we will focus instead of the syntactic interpretation as
program transformers. This results in a function:
\begin{flalign*}
\mathit{eval}_1\,:\,\ \left\{ t_1\, t_2\,:\,U\right\}
  \left\{ c_1\, c_2\,:\, t_1\,\leftrightarrow\,t_2\right\}
  (ce\,\,c_1 \Leftrightarrow\,c_2) \rightarrow (t_1\,\leftrightarrow\,t_2) \\
\end{flalign*}
This function is ``deeply dependent'': given the type of the
rewrite $ce$ to apply, both the input $c_1$ and output $c_2$ are
almost entirely determined!  Let us take for example the
second combinator in Fig.~\ref{figi}:
\[
  {(c_1 \otimes (c_2 \oplus c_3)) \odot \distl \Leftrightarrow \distl \odot ((c_1 \otimes c_2) \oplus (c_1 \otimes c_3))}
\]
\noindent which we can name \AgdaInductiveConstructor{distl⇔l}.
Interpreting this as a rewrite from the program on the left to the
one on the right requires ``pattern matching'' on the left
structure which contains $3$ arbitrary combinators, from which
we can \emph{reconstruct} the program on the right. Rewrites
such as \AgdaInductiveConstructor{distl⇔l} are one-step rewrites,
in the same way that \distl\ is a constant of the base term language
of $\Pi$.  There is one additional wrinkle. There is naturally
an opposite combinator, which interprets the above from right
to left; let us call it \AgdaInductiveConstructor{distl⇔r}. It
would appear to require \emph{non-linear pattern-matching}
since the right-hand-side contains $c_1$ twice. That is however not
the case! The definition of \AgdaInductiveConstructor{distl⇔r} has
$5$ implicit arguments, $3$ of which are $c_1, c_2$, and $c_3$,
which then completely force the ``shape'' of the overall pattern.
Thus the mere mention of \AgdaInductiveConstructor{distl⇔r} is
enough to resolve the apparent use of a non-linear pattern.
This is why \evalone\ was called ``deeply dependent''
above: once the name of the combinator is given, the rest
follows.

\newcommand{\transLR}{\AgdaInductiveConstructor{trans⇔}}
If all expressible transformations were single-step only, this
would hardly justify calling this an ``operational semantics,''
as we would hardly have a programming language. However, level-2
of $\Pi$ has combinators as well: two are in
Fig.~\ref{fige} and two are in Fig.~\ref{figh}. The most interesting
one is ``sequential composition,'' which is the middle one at the
bottom of Fig.~\ref{figh}. Since $\Leftrightarrow$ represents
an equivalence, sequential composition in this context is the
same as transitivity of equivalences, as thus we have chosen to
name this \transLR. When evaluating \transLR, we could cheat:
we know that the eventual answer must be, and we could just
return that. But this is not operational in any real sense, as
that skips over the intermediate steps. We would like to be able
to ``trace'' the rewrite. Thus the evaluation of
$\transLR~r_0~r_1$ where $r_0 : c_0 \leftrightarrow c_1$ and
$r_1 : c_1 \leftrightarrow c_2$
should apply \evalone\ to both $r_0$ and $r_1$. Furthermore,
after applying $r_0$, we should be able to witness that the result
is indeed $c_1$, so that we may continue. This last requirement
forces us to define a new function, mutually recursively with
\evalone, for this task:
\begin{flalign*}
\mathit{exact}\,:\,\ \left\{ t_1\, t_2\,:\,U\right\}
  \left\{ c_1\, c_2\,:\, t_1\,\leftrightarrow\,t_2\right\}
  (ce\,\,c_1 \Leftrightarrow\,c_2) \rightarrow \evalone\, ce \equiv c_2 \\
\end{flalign*}
\noindent If we are careful in our construction of \evalone, the
definition of $\mathit{exact}$ is quite straightforward, i.e.  almost
all cases are immediately provable by reflexivity.

This then lets us define the $\transLR~r_0~r_1$ case properly: we
first evaluate $r_0$ and get a result combinator, witness that this
result type is indeed exactly what we expect, and proceed to evaluate
$r_1$ where we specify that the $r_1$'s left-hand side must be
$\evalone\ r_0$; we can use the Agda keyword \AgdaKeyword{rewrite} to
make this match $c_2$ ``on the nose'' (otherwise the call would be
ill-typed).  This then forces us to use \AgdaKeyword{rewrite} also in
the implementation of the \transLR\ case in $\mathit{exact}$.

The other three combinators are much simpler, as simple
recursive calls are sufficient.

\subsection{Example}\label{sec:level2-example}

We can now illustrate how this all works with a small example.
Consider a circuit that takes an input type consisting of three values
\Tree [ {\small a} [ {\small b} {\small c} ] ]~
and swaps the leftmost value with the rightmost value to produce
\Tree [ {\small c} [ {\small b} {\small a} ] ]~.
We can implement two such circuits using our Agda library for $\Pi$:

\begin{code}%
\>[0]\AgdaFunction{swap{-}fl1}\AgdaSpace{}%
\AgdaFunction{swap{-}fl2}\AgdaSpace{}%
\AgdaSymbol{:}\AgdaSpace{}%
\AgdaSymbol{\{}\AgdaBound{a}\AgdaSpace{}%
\AgdaBound{b}\AgdaSpace{}%
\AgdaBound{c}\AgdaSpace{}%
\AgdaSymbol{:}\AgdaSpace{}%
\AgdaDatatype{U}\AgdaSymbol{\}}\AgdaSpace{}%
\AgdaSymbol{→}\AgdaSpace{}%
\AgdaInductiveConstructor{PLUS}\AgdaSpace{}%
\AgdaBound{a}\AgdaSpace{}%
\AgdaSymbol{(}\AgdaInductiveConstructor{PLUS}\AgdaSpace{}%
\AgdaBound{b}\AgdaSpace{}%
\AgdaBound{c}\AgdaSymbol{)}\AgdaSpace{}%
\AgdaDatatype{⟷}\AgdaSpace{}%
\AgdaInductiveConstructor{PLUS}\AgdaSpace{}%
\AgdaBound{c}\AgdaSpace{}%
\AgdaSymbol{(}\AgdaInductiveConstructor{PLUS}\AgdaSpace{}%
\AgdaBound{b}\AgdaSpace{}%
\AgdaBound{a}\AgdaSymbol{)}\<%
\\
\>[0]\AgdaFunction{swap{-}fl1}\AgdaSpace{}%
\AgdaSymbol{=}\AgdaSpace{}%
\AgdaInductiveConstructor{assocl₊}\AgdaSpace{}%
\AgdaInductiveConstructor{◎}\AgdaSpace{}%
\AgdaInductiveConstructor{swap₊}\AgdaSpace{}%
\AgdaInductiveConstructor{◎}\AgdaSpace{}%
\AgdaSymbol{(}\AgdaInductiveConstructor{id⟷}\AgdaSpace{}%
\AgdaInductiveConstructor{⊕}\AgdaSpace{}%
\AgdaInductiveConstructor{swap₊}\AgdaSymbol{)}\<%
\\
\\[\AgdaEmptyExtraSkip]%
\>[0]\AgdaFunction{swap{-}fl2}\AgdaSpace{}%
\AgdaSymbol{=}%
\>[52I]\AgdaSymbol{(}\AgdaInductiveConstructor{id⟷}\AgdaSpace{}%
\AgdaInductiveConstructor{⊕}\AgdaSpace{}%
\AgdaInductiveConstructor{swap₊}\AgdaSymbol{)}\AgdaSpace{}%
\AgdaInductiveConstructor{◎}\<%
\\
\>[.]\<[52I]%
\>[11]\AgdaInductiveConstructor{assocl₊}\AgdaSpace{}%
\AgdaInductiveConstructor{◎}\<%
\\
\>[11]\AgdaSymbol{(}\AgdaInductiveConstructor{swap₊}\AgdaSpace{}%
\AgdaInductiveConstructor{⊕}\AgdaSpace{}%
\AgdaInductiveConstructor{id⟷}\AgdaSymbol{)}\AgdaSpace{}%
\AgdaInductiveConstructor{◎}\<%
\\
\>[11]\AgdaInductiveConstructor{assocr₊}\AgdaSpace{}%
\AgdaInductiveConstructor{◎}\<%
\\
\>[11]\AgdaSymbol{(}\AgdaInductiveConstructor{id⟷}\AgdaSpace{}%
\AgdaInductiveConstructor{⊕}\AgdaSpace{}%
\AgdaInductiveConstructor{swap₊}\AgdaSymbol{)}\<%
\end{code}

\noindent The first implementation rewrites the incoming values as follows:
\[
\Tree [ {\small a} [ {\small b} {\small c} ] ] ~\to~
\Tree [ [ {\small a} {\small b} ] {\small c} ] ~\to~
\Tree [ {\small c} [ {\small a} {\small b} ] ] ~\to~
\Tree [ {\small c} [ {\small b} {\small a} ] ] ~.
\]
\noindent
The second implementation rewrites the incoming values as follows:
\[
\Tree [ {\small a} [ {\small b} {\small c} ] ] ~\to~
\Tree [ {\small a} [ {\small c} {\small b} ] ] ~\to~
\Tree [ [ {\small a} {\small c} ] {\small b} ] ~\to~
\Tree [ [ {\small c} {\small a} ] {\small b} ] ~\to~
\Tree [ {\small c} [ {\small a} {\small b} ] ] ~\to~
\Tree [ {\small c} [ {\small b} {\small a} ] ] ~.
\]
\noindent The two circuits are extensionally equal. Using the level-2
isomorphisms we can \emph{explicitly} construct a sequence of
rewriting steps that transforms the second circuit to the first.

We write such proofs in an equational style: in the left column, we have
the current combinator which is equivalent to the first one, and in
the right column, the justification for that equivalence. The
joining combinator is syntactic sugar for \transLR.  The transformation
could be written (using \transLR) by just giving all the pieces in
the right hand column --- but such transformations are very hard for
humans to understand and follow.

The proof
can be read as follows: the first three lines ``refocus'' from a right-associated
isomorphism onto the (left-associated) composition of the first $3$ isomorphisms;
then apply a complex rewrite on these (the ``hexagon'' coherence condition
of symmetric braided monoidal categories); this exposes two inverse combinators
next to each other --- so we have to refocus on these to eliminate them; we
finally re-associate to get the result.

\renewcommand{\AgdaIndentSpace}{\;\;}
\setlength\mathindent{0.5em}

\medskip
{\footnotesize
\begin{samepage}
\begin{code}%
\>[0]\AgdaFunction{swap{-}fl2⇔swap{-}fl1}\AgdaSpace{}%
\AgdaSymbol{:}\AgdaSpace{}%
\AgdaSymbol{\{}\AgdaBound{a}\AgdaSpace{}%
\AgdaBound{b}\AgdaSpace{}%
\AgdaBound{c}\AgdaSpace{}%
\AgdaSymbol{:}\AgdaSpace{}%
\AgdaDatatype{U}\AgdaSymbol{\}}\AgdaSpace{}%
\AgdaSymbol{→}\AgdaSpace{}%
\AgdaFunction{swap{-}fl2}\AgdaSpace{}%
\AgdaSymbol{\{}\AgdaBound{a}\AgdaSymbol{\}}\AgdaSpace{}%
\AgdaSymbol{\{}\AgdaBound{b}\AgdaSymbol{\}}\AgdaSpace{}%
\AgdaSymbol{\{}\AgdaBound{c}\AgdaSymbol{\}}\AgdaSpace{}%
\AgdaOperator{\AgdaDatatype{⇔}}\AgdaSpace{}%
\AgdaFunction{swap{-}fl1}\<%
\\
\>[0]\AgdaFunction{swap{-}fl2⇔swap{-}fl1}\AgdaSpace{}%
\AgdaSymbol{=}\<%
\\
\>[0][@{}l@{\AgdaIndent{0}}]%
\>[2]\AgdaSymbol{((}\AgdaInductiveConstructor{id⟷}\AgdaSpace{}%
\AgdaOperator{\AgdaInductiveConstructor{⊕}}\AgdaSpace{}%
\AgdaInductiveConstructor{swap₊}\AgdaSymbol{)}\AgdaSpace{}%
\AgdaOperator{\AgdaInductiveConstructor{◎}}\AgdaSpace{}%
\AgdaInductiveConstructor{assocl₊}\AgdaSpace{}%
\AgdaOperator{\AgdaInductiveConstructor{◎}}\AgdaSpace{}%
\AgdaSymbol{(}\AgdaInductiveConstructor{swap₊}\AgdaSpace{}%
\AgdaOperator{\AgdaInductiveConstructor{⊕}}\AgdaSpace{}%
\AgdaInductiveConstructor{id⟷}\AgdaSymbol{)}\AgdaSpace{}%
\AgdaOperator{\AgdaInductiveConstructor{◎}}\AgdaSpace{}%
\AgdaInductiveConstructor{assocr₊}\AgdaSpace{}%
\AgdaOperator{\AgdaInductiveConstructor{◎}}\AgdaSpace{}%
\AgdaSymbol{(}\AgdaInductiveConstructor{id⟷}\AgdaSpace{}%
\AgdaOperator{\AgdaInductiveConstructor{⊕}}\AgdaSpace{}%
\AgdaInductiveConstructor{swap₊}\AgdaSymbol{))}%
\>[75]\AgdaOperator{\AgdaFunction{⇔⟨}}\AgdaSpace{}%
\AgdaInductiveConstructor{id⇔}\AgdaSpace{}%
\AgdaOperator{\AgdaInductiveConstructor{⊡}}\AgdaSpace{}%
\AgdaInductiveConstructor{assoc◎l}\AgdaSpace{}%
\AgdaOperator{\AgdaFunction{⟩}}\<%
\\
\>[2]\AgdaSymbol{((}\AgdaInductiveConstructor{id⟷}\AgdaSpace{}%
\AgdaOperator{\AgdaInductiveConstructor{⊕}}\AgdaSpace{}%
\AgdaInductiveConstructor{swap₊}\AgdaSymbol{)}\AgdaSpace{}%
\AgdaOperator{\AgdaInductiveConstructor{◎}}\AgdaSpace{}%
\AgdaSymbol{(}\AgdaInductiveConstructor{assocl₊}\AgdaSpace{}%
\AgdaOperator{\AgdaInductiveConstructor{◎}}\AgdaSpace{}%
\AgdaSymbol{(}\AgdaInductiveConstructor{swap₊}\AgdaSpace{}%
\AgdaOperator{\AgdaInductiveConstructor{⊕}}\AgdaSpace{}%
\AgdaInductiveConstructor{id⟷}\AgdaSymbol{))}\AgdaSpace{}%
\AgdaOperator{\AgdaInductiveConstructor{◎}}\AgdaSpace{}%
\AgdaInductiveConstructor{assocr₊}\AgdaSpace{}%
\AgdaOperator{\AgdaInductiveConstructor{◎}}\AgdaSpace{}%
\AgdaSymbol{(}\AgdaInductiveConstructor{id⟷}\AgdaSpace{}%
\AgdaOperator{\AgdaInductiveConstructor{⊕}}\AgdaSpace{}%
\AgdaInductiveConstructor{swap₊}\AgdaSymbol{))}%
\>[75]\AgdaOperator{\AgdaFunction{⇔⟨}}\AgdaSpace{}%
\AgdaInductiveConstructor{assoc◎l}\AgdaSpace{}%
\AgdaOperator{\AgdaFunction{⟩}}\<%
\\
\>[2]\AgdaSymbol{(((}\AgdaInductiveConstructor{id⟷}\AgdaSpace{}%
\AgdaOperator{\AgdaInductiveConstructor{⊕}}\AgdaSpace{}%
\AgdaInductiveConstructor{swap₊}\AgdaSymbol{)}\AgdaSpace{}%
\AgdaOperator{\AgdaInductiveConstructor{◎}}\AgdaSpace{}%
\AgdaInductiveConstructor{assocl₊}\AgdaSpace{}%
\AgdaOperator{\AgdaInductiveConstructor{◎}}\AgdaSpace{}%
\AgdaSymbol{(}\AgdaInductiveConstructor{swap₊}\AgdaSpace{}%
\AgdaOperator{\AgdaInductiveConstructor{⊕}}\AgdaSpace{}%
\AgdaInductiveConstructor{id⟷}\AgdaSymbol{))}\AgdaSpace{}%
\AgdaOperator{\AgdaInductiveConstructor{◎}}\AgdaSpace{}%
\AgdaInductiveConstructor{assocr₊}\AgdaSpace{}%
\AgdaOperator{\AgdaInductiveConstructor{◎}}\AgdaSpace{}%
\AgdaSymbol{(}\AgdaInductiveConstructor{id⟷}\AgdaSpace{}%
\AgdaOperator{\AgdaInductiveConstructor{⊕}}\AgdaSpace{}%
\AgdaInductiveConstructor{swap₊}\AgdaSymbol{))}%
\>[75]\AgdaOperator{\AgdaFunction{⇔⟨}}\AgdaSpace{}%
\AgdaInductiveConstructor{assoc◎l}\AgdaSpace{}%
\AgdaOperator{\AgdaInductiveConstructor{⊡}}\AgdaSpace{}%
\AgdaInductiveConstructor{id⇔}\AgdaSpace{}%
\AgdaOperator{\AgdaFunction{⟩}}\<%
\\
\>[2]\AgdaSymbol{((((}\AgdaInductiveConstructor{id⟷}\AgdaSpace{}%
\AgdaOperator{\AgdaInductiveConstructor{⊕}}\AgdaSpace{}%
\AgdaInductiveConstructor{swap₊}\AgdaSymbol{)}\AgdaSpace{}%
\AgdaOperator{\AgdaInductiveConstructor{◎}}\AgdaSpace{}%
\AgdaInductiveConstructor{assocl₊}\AgdaSymbol{)}\AgdaSpace{}%
\AgdaOperator{\AgdaInductiveConstructor{◎}}\AgdaSpace{}%
\AgdaSymbol{(}\AgdaInductiveConstructor{swap₊}\AgdaSpace{}%
\AgdaOperator{\AgdaInductiveConstructor{⊕}}\AgdaSpace{}%
\AgdaInductiveConstructor{id⟷}\AgdaSymbol{))}\AgdaSpace{}%
\AgdaOperator{\AgdaInductiveConstructor{◎}}\AgdaSpace{}%
\AgdaInductiveConstructor{assocr₊}\AgdaSpace{}%
\AgdaOperator{\AgdaInductiveConstructor{◎}}\AgdaSpace{}%
\AgdaSymbol{(}\AgdaInductiveConstructor{id⟷}\AgdaSpace{}%
\AgdaOperator{\AgdaInductiveConstructor{⊕}}\AgdaSpace{}%
\AgdaInductiveConstructor{swap₊}\AgdaSymbol{))}%
\>[75]\AgdaOperator{\AgdaFunction{⇔⟨}}\AgdaSpace{}%
\AgdaInductiveConstructor{hexagonl⊕r}\AgdaSpace{}%
\AgdaOperator{\AgdaInductiveConstructor{⊡}}\AgdaSpace{}%
\AgdaInductiveConstructor{id⇔}\AgdaSpace{}%
\AgdaOperator{\AgdaFunction{⟩}}\<%
\\
\>[2]\AgdaSymbol{(((}\AgdaInductiveConstructor{assocl₊}\AgdaSpace{}%
\AgdaOperator{\AgdaInductiveConstructor{◎}}\AgdaSpace{}%
\AgdaInductiveConstructor{swap₊}\AgdaSymbol{)}\AgdaSpace{}%
\AgdaOperator{\AgdaInductiveConstructor{◎}}\AgdaSpace{}%
\AgdaInductiveConstructor{assocl₊}\AgdaSymbol{)}\AgdaSpace{}%
\AgdaOperator{\AgdaInductiveConstructor{◎}}\AgdaSpace{}%
\AgdaInductiveConstructor{assocr₊}%
\>[44]\AgdaOperator{\AgdaInductiveConstructor{◎}}\AgdaSpace{}%
\AgdaSymbol{(}\AgdaInductiveConstructor{id⟷}\AgdaSpace{}%
\AgdaOperator{\AgdaInductiveConstructor{⊕}}\AgdaSpace{}%
\AgdaInductiveConstructor{swap₊}\AgdaSymbol{))}%
\>[75]\AgdaOperator{\AgdaFunction{⇔⟨}}\AgdaSpace{}%
\AgdaInductiveConstructor{assoc◎r}\AgdaSpace{}%
\AgdaOperator{\AgdaFunction{⟩}}\<%
\\
\>[2]\AgdaSymbol{((}\AgdaInductiveConstructor{assocl₊}\AgdaSpace{}%
\AgdaOperator{\AgdaInductiveConstructor{◎}}\AgdaSpace{}%
\AgdaInductiveConstructor{swap₊}\AgdaSymbol{)}\AgdaSpace{}%
\AgdaOperator{\AgdaInductiveConstructor{◎}}\AgdaSpace{}%
\AgdaInductiveConstructor{assocl₊}\AgdaSpace{}%
\AgdaOperator{\AgdaInductiveConstructor{◎}}\AgdaSpace{}%
\AgdaInductiveConstructor{assocr₊}%
\>[42]\AgdaOperator{\AgdaInductiveConstructor{◎}}\AgdaSpace{}%
\AgdaSymbol{(}\AgdaInductiveConstructor{id⟷}\AgdaSpace{}%
\AgdaOperator{\AgdaInductiveConstructor{⊕}}\AgdaSpace{}%
\AgdaInductiveConstructor{swap₊}\AgdaSymbol{))}%
\>[75]\AgdaOperator{\AgdaFunction{⇔⟨}}\AgdaSpace{}%
\AgdaInductiveConstructor{id⇔}\AgdaSpace{}%
\AgdaOperator{\AgdaInductiveConstructor{⊡}}\AgdaSpace{}%
\AgdaInductiveConstructor{assoc◎l}\AgdaSpace{}%
\AgdaOperator{\AgdaFunction{⟩}}\<%
\\
\>[2]\AgdaSymbol{((}\AgdaInductiveConstructor{assocl₊}\AgdaSpace{}%
\AgdaOperator{\AgdaInductiveConstructor{◎}}\AgdaSpace{}%
\AgdaInductiveConstructor{swap₊}\AgdaSymbol{)}\AgdaSpace{}%
\AgdaOperator{\AgdaInductiveConstructor{◎}}\AgdaSpace{}%
\AgdaSymbol{(}\AgdaInductiveConstructor{assocl₊}\AgdaSpace{}%
\AgdaOperator{\AgdaInductiveConstructor{◎}}\AgdaSpace{}%
\AgdaInductiveConstructor{assocr₊}\AgdaSymbol{)}%
\>[44]\AgdaOperator{\AgdaInductiveConstructor{◎}}\AgdaSpace{}%
\AgdaSymbol{(}\AgdaInductiveConstructor{id⟷}\AgdaSpace{}%
\AgdaOperator{\AgdaInductiveConstructor{⊕}}\AgdaSpace{}%
\AgdaInductiveConstructor{swap₊}\AgdaSymbol{))}%
\>[75]\AgdaOperator{\AgdaFunction{⇔⟨}}\AgdaSpace{}%
\AgdaInductiveConstructor{id⇔}\AgdaSpace{}%
\AgdaOperator{\AgdaInductiveConstructor{⊡}}\AgdaSpace{}%
\AgdaSymbol{(}\AgdaInductiveConstructor{linv◎l}\AgdaSpace{}%
\AgdaOperator{\AgdaInductiveConstructor{⊡}}\AgdaSpace{}%
\AgdaInductiveConstructor{id⇔}\AgdaSymbol{)}\AgdaSpace{}%
\AgdaOperator{\AgdaFunction{⟩}}\<%
\\
\>[2]\AgdaSymbol{((}\AgdaInductiveConstructor{assocl₊}\AgdaSpace{}%
\AgdaOperator{\AgdaInductiveConstructor{◎}}\AgdaSpace{}%
\AgdaInductiveConstructor{swap₊}\AgdaSymbol{)}\AgdaSpace{}%
\AgdaOperator{\AgdaInductiveConstructor{◎}}\AgdaSpace{}%
\AgdaInductiveConstructor{id⟷}%
\>[28]\AgdaOperator{\AgdaInductiveConstructor{◎}}\AgdaSpace{}%
\AgdaSymbol{(}\AgdaInductiveConstructor{id⟷}\AgdaSpace{}%
\AgdaOperator{\AgdaInductiveConstructor{⊕}}\AgdaSpace{}%
\AgdaInductiveConstructor{swap₊}\AgdaSymbol{))}%
\>[75]\AgdaOperator{\AgdaFunction{⇔⟨}}\AgdaSpace{}%
\AgdaInductiveConstructor{id⇔}\AgdaSpace{}%
\AgdaOperator{\AgdaInductiveConstructor{⊡}}\AgdaSpace{}%
\AgdaInductiveConstructor{idl◎l}\AgdaSpace{}%
\AgdaOperator{\AgdaFunction{⟩}}\<%
\\
\>[2]\AgdaSymbol{((}\AgdaInductiveConstructor{assocl₊}\AgdaSpace{}%
\AgdaOperator{\AgdaInductiveConstructor{◎}}\AgdaSpace{}%
\AgdaInductiveConstructor{swap₊}\AgdaSymbol{)}%
\>[22]\AgdaOperator{\AgdaInductiveConstructor{◎}}\AgdaSpace{}%
\AgdaSymbol{(}\AgdaInductiveConstructor{id⟷}\AgdaSpace{}%
\AgdaOperator{\AgdaInductiveConstructor{⊕}}\AgdaSpace{}%
\AgdaInductiveConstructor{swap₊}\AgdaSymbol{))}%
\>[75]\AgdaOperator{\AgdaFunction{⇔⟨}}\AgdaSpace{}%
\AgdaInductiveConstructor{assoc◎r}\AgdaSpace{}%
\AgdaOperator{\AgdaFunction{⟩}}\<%
\\
\>[2]\AgdaSymbol{((}\AgdaInductiveConstructor{assocl₊}\AgdaSpace{}%
\AgdaOperator{\AgdaInductiveConstructor{◎}}\AgdaSpace{}%
\AgdaInductiveConstructor{swap₊}\AgdaSpace{}%
\AgdaOperator{\AgdaInductiveConstructor{◎}}\AgdaSpace{}%
\AgdaSymbol{(}\AgdaInductiveConstructor{id⟷}\AgdaSpace{}%
\AgdaOperator{\AgdaInductiveConstructor{⊕}}\AgdaSpace{}%
\AgdaInductiveConstructor{swap₊}\AgdaSymbol{))}\AgdaSpace{}%
\AgdaOperator{\AgdaFunction{▤}}\AgdaSymbol{)}\<%
\end{code}
\end{samepage}
}

\renewcommand{\AgdaIndentSpace}{\AgdaSpace{}$\;\;$}

\subsection{Internal Language}

Recalling that the $\lambda$-calculus arises as the internal language
of Cartesian Closed Categories (Elliott~\cite{Elliott-2017} gives a particularly
readable account of this), we can think of $\Pi$ in similar terms, but
for symmetric Rig Groupoids instead. For example, we can ask what does
the derivation in Sec.~\ref{sec:level2-example} represent? It is
actually a ``linear'' representation of a 2-categorial commutative
diagram! In fact, it is a painfully verbose version thereof, as it
includes many \emph{refocusing} steps because our language does not
build associativity into its syntax. Categorical diagrams usually do.
Thus if we rewrite the example in diagrammatic form, eliding all uses
of associativity, but keeping explicit uses of identity transformations,
we get that \AgdaFunction{swap{-}fl2⇔swap{-}fl1} represents

\newcommand{\idd}{\mathit{id}\leftrightarrow}
\newcommand{\idf}{\mathit{id}\Leftrightarrow}
\vspace*{3mm}
\begin{tikzcd}[column sep=normal, row sep=normal]
 && (a+c)+b \arrow [r, "\swapp \oplus\idd", ""{name=U, below}] & (c+a)+b \arrow [dr, "\assocrp"] && \\
 & a+(c+b) \arrow [ur, "\assoclp"] & & & c+(a+b) \arrow [dr, "\idd\oplus\swapp"] &  \\
a+(b+c) \arrow [ur, "\idd\oplus\swapp"] \arrow [r, "\assoclp"]
  \arrow [dr, "\assoclp"]
  \arrow [ddr, swap, "\assoclp"]
    & (a+b)+c \arrow [r, "\swapp"] &
    c+(a+b) \arrow [r, swap, "\assoclp", ""{name=D, above}]
    & |[alias=Z]| (c+a)+b \arrow [r, "\assocrp"] &c+(a+b) \arrow [r, "\idd\oplus\swapp"] & c+(b+a) \\
 & (a+b)+c \arrow [dr, "\swapp"] &&&& \\
 & (a+b)+c \arrow [dr, swap, "\swapp"] & c+(a+b) \arrow [rr, swap, "\idd", ""{name=DD, above}]
             \arrow [d, Rightarrow, "\idf\, \mathit{idl}\odot{l}"] &&
    c+(a+b) \arrow [ruu, "\idd\oplus\swapp"] & \\
 && c+(a+b) \arrow [rrruuu, bend right = 40, swap, "\idd\oplus\swapp"] && \\
 \arrow[Rightarrow, from=U, to=D, "\mathit{hexagon}\oplus{r}\, \boxdot\, \idf"]
 \arrow[Rightarrow, from=Z, to=DD, swap, "\idf\boxdot\mathit{linv}\odot{l}\,\boxdot\,\idf"]
\end{tikzcd}

\noindent For some, the above diagram will be clearer --- it is only three layers
high rather than nine! Others will prefer the more programmatic feel of the
original definition.

We would be remiss in letting the reader believe that the above is ``the''
categorical diagram that would be found in categorical textbooks. Rather,
congruence would be used to elide the $\idf$. Furthermore, the various arrows
would also be named differently --- our \assoclp\ is often named $\alpha$,
\assocrp\ is $\alpha^{-1}$, $\swapp$ is $B$ (always with subscripts).
And the two steps needed to remove inverses (i.e. first cancelling
inverse arrows, then removing the resulting identity arrow ``in context'')
are often combined into one. Here we'll simply name this operation
$\mathit{cancel}$, which could be programmed as a defined function over
$\Pi$ level-2.  The result would then be the much simpler

\vspace*{3mm}
\begin{tikzcd}[column sep=normal, row sep=normal]
 & a+(c+b) \arrow [r, "\assoclp", ""{name=U, below}] & (a+c)+b \arrow [rd, "\swapp\oplus\idd"] & & & \\
a+(b+c) \arrow [ur, "\idd\oplus\swapp"] \arrow [dr, "\assoclp"]
  & & & (c+a)+b \arrow [r, "\assocrp", ""{name=UU, below}] & c+(a+b) \arrow [r, "\idd\oplus\swapp"] & c+(b+a) \\
 & (a+b)+c \arrow [r, swap, "\swapp", ""{name=D, above}] & c+(a+b) \arrow [ur, "\assoclp"]
  \arrow [urrr, swap, "\idd\oplus\swapp", ""{name=DD,above}] & & & & \\
 \arrow[Rightarrow, from=U, to=D, "\mathit{hexagon}\oplus{r}"]
 \arrow[Rightarrow, from=UU, to=DD, "\mathit{cancel}"]
\end{tikzcd}

In other words, each (non-refocusing) line of the proof of
\AgdaFunction{swap{-}fl2⇔swap{-}fl1}\; is a complete path
from left to right in each diagram above, and the annotation
on the right-hand-side becomes the natural transformation (denoted
by vertical $\Rightarrow$) justifying the move to the next line.
The first diagram uses lines $1,4,7,8$ in full; the second
diagram collapses $7$ and $8$ into one, as well as not duplicating
parts which are related by $\idf$.

\section{Further Thoughts and Conclusions}

We conclude with a collection of open problems and avenues for further research.

\subsection{Richer Data: Infinite Sets and Topological Spaces}

The three languages we discussed only deal with the finite spaces
built from 0, 1, sums, and products. Programming practice, logic, and
mathematics all deal with richer spaces including inductive types
(e.g., the natural numbers, sequences, and trees), functions, and
graphs. Extending $\Pi$ to such domains is possible but only after one
refines the notions of reversibility and conservation of
information. One approach is to use \emph{partial isomorphisms} that
may be undefined on such inputs~\cite{infeffects,rc2011}. Another more
speculative approach is to build such spaces, topologically, based on
novel type constructions such as negative, fractional, or even
imaginary types~\cite{seventrees,roshan-thesis}.

\subsection{Information Effects}

A computational model that enforces the principle of conservation of
information is arguably \emph{richer} than a conventional model that
cannot even express the notion of information. Practically the
conventional model is easily recovered by simply adding constructs
that intentionally and explicitly create or erase information. Such
constructs allow one to recover the classical perspective with the
added advantage that it is possible to reason about such creation and
erasure of information using type and effect systems, monads, or
arrows~\cite{infeffects,inversearrows}.

An interesting application of such an idea is in the field of
\emph{information-flow security}. To make this idea concrete, consider
a tiny 2-bit password = \verb|"10"| and the associated password checker:

\begin{verbatim}
check-password (guess) =
  guess == "10"
\end{verbatim}

One can ask how much information is leaked by this program assuming
the attacker has no prior knowledge except that the password is 2
bits, i.e., the four possible 2-bits are equally likely. If the
attacker guesses \verb|"10"| (with probability $1/4$) the password (2
bits) is leaked. If the attacker guesses one of the other choices
(with probability $3/4$) the number of possibilities is reduced from 4
to 3, i.e., the attacker learns $\log{4} - \log{3}$ bits of
information. So in general the attacker learns:
\[\begin{array}{ll}
   &  1/4 * 2 + 3/4 (\log{4} - \log{3}) \\
  =&  1/4 \log{4} + 3/4 \log{4/3} \\
  =&  - 1/4 \log{1/4} - 3/4 \log{3/4} \\
  \sim& 0.8 \mbox{~bits~in~the~first~probe}
\end{array}\]
This is a significant amount of information. But of course this is
only because the password is so short: if the password was 8
restricted ASCII characters (6 bits), the attacker would only learn
0.00001 bits in the first probe.

An alternative formulation of the problem is to view the input as a
random variable with 4 possibilities and a uniform distribution (i.e.,
with 2 bits of information) and the output as another random variable
with 4 possibilities but with the distribution
$\{ (\mathit{True}, 1/4), (\mathit{False}, 3/4) \}$ which contains 0.8 bits of
information. Thus 2 input bits of information were given to the
password checker and only 0.8 were produced. Where did the 1.2 bits of
information go? By the Landauer Principle, these 1.2 bits must be
accounted by an \emph{implicit} \emph{erasure} operation in the
program. By writing the password checker in an extension of $\Pi$, the
erasure construct becomes explicit and the information leak becomes
exposed in the syntactic structure of the program~\cite{infeffects}.

\subsection{Theseus and Quantum Control}

The $\Pi$ family of languages semantically captures the principles of
reversibility and conservation of information. As a programming
language it has some mixed properties: small programs are relatively
easy to write; for some special classes of programs, it is even
possible to define a methodology to write large $\Pi$ programs,
including a meta-circular interpreter for $\Pi$~\cite{isoint}. In
general, however, the point-free style of combinators used in $\Pi$
becomes awkward and a new approach appears more suitable. To that end,
we note that $\Pi$ encodes the most elementary control structure in a
programming language-- which is the ability to conditionally execute
one of several possible code fragments-- using combinators. Expressing
such an abstraction using combinators or even predicates and nested
\textbf{if}-expressions makes it difficult for both humans and
compilers to write, understand, and reason about the control flow
structure of the program. Instead, in modern functional languages,
this control flow paradigm is elegantly expressed using
\emph{pattern-matching}. This approach yields code that is not only
more concise and readable but also enables the compiler to easily
verify two crucial properties: (i) non-overlapping patterns and (ii)
exhaustive coverage of a datatype using a collection of
patterns. Indeed most compilers for functional languages perform these
checks, warning the user when they are violated. At a more fundamental
level, e.g., in type theories and proof assistants, these properties
are actually necessary for correct reasoning about programs. Our
insight is that these properties, perhaps surprisingly, are sufficient
to produce a simple and intuitive first-order reversible programming
language which we call \emph{Theseus}.

\begin{figure}[t]
  \centering
  \begin{minipage}{0.4\linewidth}
    \begin{verbatim}
f :: Either Int Int -> a
f (Left 0)     = undefined
f (Left (n+1)) = undefined
f (Right n)    = undefined
\end{verbatim}\vspace{-4ex}
    \caption{A skeleton}\label{fig:intro-ex-1}
  \end{minipage}
  \hfill
  \begin{minipage}{0.4\linewidth}
\begin{verbatim}
g :: (Bool,Int) -> a
g (False,n)  = undefined
g (True,0)   = undefined
g (True,n+1) = undefined
\end{verbatim}\vspace{-4ex}
    \caption{Another skeleton}\label{fig:intro-ex-2}
  \end{minipage}
  \\[3ex]
  \begin{minipage}{0.6\linewidth}
\begin{verbatim}
h :: Either Int Int <-> (Bool,Int)
h (Left 0)     = (True,0)
h (Left (n+1)) = (False,n)
h (Right n)    = (True,n+1)
\end{verbatim}\vspace{-4ex}
    \caption{An isomorphism}\label{fig:intro-ex-3}
  \end{minipage}
\end{figure}

We provide a small illustrative example, written in a Haskell-like
syntax.  Fig.~\ref{fig:intro-ex-1} gives the skeleton of a function
\verb|f| that accepts a value of type \verb|Either Int| \verb|Int|;
the patterns on the left-hand side exhaustively cover every possible
incoming value and are non-overlapping. Similarly,
Fig.~\ref{fig:intro-ex-2} gives the skeleton for a function~\verb|g|
that accepts a value of type \verb|(Bool,Int)|; again the patterns on
the left-hand side exhaustively cover every possible incoming value
and are non-overlapping. Now we claim that since the types
\verb|Either Int Int| and \verb|(Bool,Int)| are isomorphic, we can
combine the patterns of \verb|f| and \verb|g| into \emph{symmetric
  pattern-matching clauses} to produce a reversible function between
the types \verb|Either Int Int| and
\verb|(Bool,Int)|. Fig.~\ref{fig:intro-ex-3} gives one such function;
there, we suggestively use \verb|<->| to indicate that the function
can be executed in either direction. This reversible function is
obtained by simply combining the non-overlapping exhaustive patterns
on the two sides of a clause. In order to be well-formed in either
direction, these clauses are subject to the constraint that each
variable occurring on one side must occur exactly once on the other
side (and with the same type). Thus it is acceptable to swap the
second and third right-hand sides of \verb|h| but not the first and
second ones. With some additional work, it is possible to extend
Theseus to a full-fledged reversible programming
language~\cite{theseus}. With just one additional insight, Theseus can
be extended with superpositions and becomes a quantum programming
language~\cite{10.1007/978-3-319-89366-2_19}.

\subsection{Quantum Speed-up}

A rather remarkable but somehow overlooked paper is
``Quantum speedup and Categorical Distributivity'' by
Peter Hines~\cite{hines2013quantum}. Here he shows that
the heart of Shor's algorithm can be reduced to an
operation $!^N()$\ (expressible in $\Pi$), which can
be expressed, via a factorization, in an exponentially
faster manner. The key to this efficient factorization
is exactly the coherence conditions of Laplaza~\cite{laplaza72},
which also feature prominently in our work. Proving his
key Lemma 2 in $\Pi$ could be quite instructive in revealing
which level-2 combinators are crucial for this result.

\subsection{Summary}

The entire edifice of computer science including its mainstream models
of computations, programming languages, and logics is founded on
\emph{classical physics.} While much of the world phenomena can be
approximated with classical physics, we are reaching a revolutionary
period of quantum technology that challenges many of the classical
assumptions. It remains to be seen how computer science will adapt to
this quantum revolution but we believe that additional physical
principles inspired by quantum mechanics will have to be embraced in
our computational thinking. This paper focused on one such principle
--- \emph{conservation of information} --- and explored some of its
exciting implications to the field of computer science.

\section*{Acknowledgements} We would like to thank the numerous
students and colleagues who participated in various aspects of this
research and who provided valuable feedback and constructive criticism.

\bibliographystyle{unsrt}
\bibliography{cites}
\end{document}